\documentclass[aps,prc,reprint,superscriptaddress,floatfix]{revtex4-1} 
\usepackage{graphicx}
\usepackage{dcolumn}
\usepackage{amsmath} 
\usepackage{physics}
\usepackage{subfloat}
\usepackage{color}
\usepackage{bm}
\usepackage{threeparttable}

\newcommand{\nuc}[2]{$^{#1}$#2}

\newcommand{\zps}{$0^{+}$ states}

\newcommand{\dpsrf}[2]{$d(^{#1}$Sr$,p)^{#2}$Sr}
\newcommand{\dtsr}[1]{$d(^{#1}$Sr$,t)$}
\newcommand{\dtsrf}[2]{$d(^{#1}$Sr$,t)^{#2}$Sr}
\newcommand{\ppsr}[1]{$p(^{#1}$Sr$,p)$}
\newcommand{\ddsr}[1]{$d(^{#1}$Sr$,d)$}

\newcommand{\sfacs}{spectroscopic factors}
\newcommand\T{\rule{0pt}{2.6ex}}       
\newcommand\B{\rule[-1.2ex]{0pt}{0pt}} 

\graphicspath{{./}}
\begin{document}

\title{Single-particle structure in neutron-rich Sr isotopes approaching the \boldmath{$N=60$} shape transition }
\author{S.~Cruz}
\affiliation{Department of Physics and Astronomy, University of British Columbia, Vancouver, BC V6T 1Z4, Canada}
\affiliation{TRIUMF - Canada's Particle Accelerator Centre, Vancouver, BC V6T 2A3, Canada}
\author{K.~Wimmer}
\email{Corresponding author: k.wimmer@csic.es}
\affiliation{Instituto de Estructura de la Materia, CSIC, E-28006 Madrid, Spain}
\affiliation{Department of Physics, Central Michigan University, Mt Pleasant, MI 48859, USA}
\affiliation{Department of Physics, The University of Tokyo, Hongo, Bunkyo-ku, Tokyo 113-0033, Japan}
\author{S. S. Bhattacharjee}
\affiliation{TRIUMF - Canada's Particle Accelerator Centre, Vancouver, BC V6T 2A3, Canada}
\author{P.C.~Bender}
\affiliation{TRIUMF - Canada's Particle Accelerator Centre, Vancouver, BC V6T 2A3, Canada}
\author{G.~Hackman}
\affiliation{TRIUMF - Canada's Particle Accelerator Centre, Vancouver, BC V6T 2A3, Canada}
\author{R.~Kr\"{u}cken}
\affiliation{Department of Physics and Astronomy, University of British Columbia, Vancouver, BC V6T 1Z4, Canada}
\affiliation{TRIUMF - Canada's Particle Accelerator Centre, Vancouver, BC V6T 2A3, Canada}
\author{F.~Ames}
\affiliation{TRIUMF - Canada's Particle Accelerator Centre, Vancouver, BC V6T 2A3, Canada}
\affiliation{Department of Astronomy and Physics, Saint Mary's University, Halifax, NS B3H 3C2, Canada}
\author{C.~Andreoiu}
\affiliation{Department of Chemistry, Simon Fraser University, Burnaby, BC V5A 1S6, Canada}
\author{R.A.E.~Austin}
\affiliation{Department of Astronomy and Physics, Saint Mary's University, Halifax, NS B3H 3C2, Canada}
\author{C.S.~Bancroft}
\affiliation{Department of Physics, Central Michigan University, Mt Pleasant, MI 48859, USA}
\author{R.~Braid}
\affiliation{Department of Physics, Colorado School of Mines, Golden, CO 80401, USA}
\author{T.~Bruhn}
\affiliation{TRIUMF - Canada's Particle Accelerator Centre, Vancouver, BC V6T 2A3, Canada}
\author{W.N.~Catford}
\affiliation{Department of Physics, University of Surrey, Guildford, Surrey, GU2 7XH, United Kingdom}
\author{A.~Cheeseman}
\affiliation{TRIUMF - Canada's Particle Accelerator Centre, Vancouver, BC V6T 2A3, Canada}
\author{A.~Chester}
\affiliation{Department of Physics, Simon Fraser University, Burnaby, BC V5A 1S6, Canada}
\author{D.S.~Cross}
\affiliation{Department of Chemistry, Simon Fraser University, Burnaby, BC V5A 1S6, Canada}
\author{C.Aa.~Diget}
\affiliation{Department of Physics, University of York, York, YO10 5DD, United Kingdom}
\author{T.~Drake}
\affiliation{Department of Physics, University of Toronto, Toronto, ON M5S 1A7, Canada}
\author{A.B.~Garnsworthy}
\affiliation{TRIUMF - Canada's Particle Accelerator Centre, Vancouver, BC V6T 2A3, Canada}
\author{R.~Kanungo}
\affiliation{TRIUMF - Canada's Particle Accelerator Centre, Vancouver, BC V6T 2A3, Canada}
\affiliation{Department of Astronomy and Physics, Saint Mary's University, Halifax, NS B3H 3C2, Canada}
\author{A.~Knapton}
\affiliation{Department of Physics, University of Surrey, Guildford, Surrey, GU2 7XH, United Kingdom}
\author{W.~Korten}
\affiliation{IRFU, CEA, Universit\'{e} Paris-Saclay, F-91191 Gif-sur-Yvette, France}
\affiliation{TRIUMF - Canada's Particle Accelerator Centre, Vancouver, BC V6T 2A3, Canada}
\author{K.~Kuhn}
\affiliation{Department of Physics, Colorado School of Mines, Golden, CO 80401, USA}
\author{J.~Lassen}
\affiliation{TRIUMF - Canada's Particle Accelerator Centre, Vancouver, BC V6T 2A3, Canada}
\affiliation{Department of Physics, Simon Fraser University, Burnaby, BC V5A 1S6, Canada}
\author{R.~Laxdal}
\author{M.~Marchetto}
\affiliation{TRIUMF - Canada's Particle Accelerator Centre, Vancouver, BC V6T 2A3, Canada}
\author{A.~Matta}
\affiliation{Department of Physics, University of Surrey, Guildford, Surrey, GU2 7XH, United Kingdom}
\affiliation{LPC, ENSICAEN, CNRS/IN2P3, UNICAEN, Normandie Universit\'{e}, 14050 Caen cedex, France}
\author{D.~Miller}
\author{M.~Moukaddam}
\affiliation{TRIUMF - Canada's Particle Accelerator Centre, Vancouver, BC V6T 2A3, Canada}
\author{N.A.~Orr}
\affiliation{LPC, ENSICAEN, CNRS/IN2P3, UNICAEN, Normandie Universit\'{e}, 14050 Caen cedex, France}
\author{N.~Sachmpazidi}
\affiliation{Department of Physics, Central Michigan University, Mt Pleasant, MI 48859, USA}
\author{A.~Sanetullaev}
\affiliation{Department of Astronomy and Physics, Saint Mary's University, Halifax, NS B3H 3C2, Canada}
\affiliation{TRIUMF - Canada's Particle Accelerator Centre, Vancouver, BC V6T 2A3, Canada}
\author{C.E.~Svensson}
\affiliation{Department of Physics, University of Guelph, Guelph, ON, N1G 2W1, Canada}
\author{N.~Terpstra}
\affiliation{Department of Physics, Central Michigan University, Mt Pleasant, MI 48859, USA}
\author{C.~Unsworth}
\affiliation{TRIUMF - Canada's Particle Accelerator Centre, Vancouver, BC V6T 2A3, Canada}
\author{P.J.~Voss}
\affiliation{Department of Chemistry, Simon Fraser University, Burnaby, BC V5A 1S6, Canada}

\date{\today}

\begin{abstract}
\begin{description}
\item[Background] Neutron-rich nuclei around neutron number $N=60$ show a dramatic shape transition from spherical ground states to prolate deformation in \nuc{98}{Sr} and heavier nuclei. 
\item[Purpose] The purpose of this study is to investigate the single-particle structure approaching the shape transitional region.  
\item[Method] The level structures of neutron-rich \nuc{93,94,95}{Sr} were studied via the \dtsr{94,95,96} one-neutron stripping reactions at TRIUMF using a beam energy of 5.5~$A$MeV. 
$\gamma$-rays emitted from excited states and recoiling charged particles were detected by using the TIGRESS and SHARC arrays, respectively. States were identified by gating on the excitation energy and, if possible, the coincident $\gamma$ radiation.
\item[Results] Triton angular distributions for the reactions populating states in ejectile nuclei \nuc{93,94,95}{Sr} were compared with distorted wave Born approximation calculations to assign and revise spin and parity quantum numbers and extract spectroscopic factors. The results were compared with shell model calculations and the reverse $(d,p)$ reactions and good agreement was obtained.
\item[Conclusions] The results for the \dtsrf{94}{93} and \dtsrf{95}{94} reactions are in good agreement with shell model calculations. A two level mixing analysis for the $0^+$ states in \nuc{94}{Sr} suggest strong mixing of two shapes. For the \dtsrf{96}{95} reaction the agreement with the shell model is less good. The configuration of the ground state of \nuc{96}{Sr} is already more complex than predicted, and therefore indications for the shape transition can already be observed before $N=60$.
\end{description}
\end{abstract}

\pacs{23.20.Lv,21.10.Hw,21.60.Cs}
\maketitle

\section{Introduction} 
The shape of a nucleus is governed by the interplay of stabilizing shell gaps which prefer a spherical shape and quadrupole correlations which tend to deform the nucleus. In the neutron-rich zirconium ($Z=40$) and strontium ($Z=38$) isotopes a sharp transition between spherical ground states for nuclei with neutron number $N\leq 59$  and strongly deformed ground states at $N=60$ and beyond has been inferred from isotope shift measurements~\cite{buchinger90}. At the same location the excitation energy of the first $2^+$ state in the even-even nuclei indicates a rapid onset of collectivity~\cite{wollnik77}.  This sudden drop in $E(2^+_1)$ at $N=60$ has been observed in Zr and Sr nuclei, but not in the heavier Mo or lighter Kr isotones. 
The origin of the shape transition has been attributed to the specifics of the proton-neutron interaction. Filling of the neutron ($\nu$) $g_{7/2}$ orbital leads to a lowering of the $g_{9/2}$ proton ($\pi$) orbital enabling proton excitations into the $\pi g_{9/2}$ orbital. The increase in occupation of the $\pi g_{9/2}$ orbital enables further promotion of particles into the $\nu g_{7/2}$ orbital leading to deformation~\cite{federman79}. Monte-Carlo shell model (MCSM) calculations in a very large model space were able to trace the effective single-particle energies of the relevant neutron orbitals for excited states~\cite{togashi16}. The results show that the structure of the $N=58$ nucleus \nuc{98}{Zr} in the excited $0^+$ state is dominated by many particle excitations to the proton $g_{9/2}$ orbital and resembles the ground state of \nuc{100}{Zr} at $N=60$. This suggests a coexistence and inversion of spherical and deformed structures at $N=60$.

The sharp drop of the $2^+$ energy is accompanied by a drop in energy of the first excited $0^+$ ($E(0^+_2)$) state. The $E(0^+_2)$ energy continuously decreases with increasing neutron number from above 3~MeV at $N = 50$ down to 1229~keV in \nuc{96}{Sr} ($N = 58$). At $N=60$ the two structures cross, the deformed $0^+$ state becomes the ground state, while the spherical one is located at only $E(0^+_2) = 215$~keV.
The interpretation that the ground state of the $N=60$ nucleus \nuc{98}{Sr} is strongly deformed while the $0^+_2$ state is spherical was recently confirmed by a measurement of the electric quadrupole moments obtained from Coulomb excitation~\cite{clement16,*clement16a}.
Interestingly, there is a third low-lying $0^+$ state present in the Sr nuclei which decreases also in energy, but not as dramatically as the deformed configuration. In \nuc{96}{Sr} the excited $0^+$ states are connected by a very strong electromagnetic transition matrix element~\cite{jungphd}. A measurement of the $(d,p)$ transfer reaction cross section from the ground state of \nuc{95}{Sr} to the $0^+$ states of \nuc{96}{Sr} showed that the excited $0^+$ states can be understood resulting from the strong mixing of two configurations with very different deformation~\cite{cruz18}. 

In this paper we investigate the single-particle structure of the neutron-rich \nuc{93-95}{Sr} nuclei by $(d,t)$ neutron removal reactions. Spin and parity assignments in the odd-$A$ \nuc{93,95}{Sr} are constrained and spectroscopic factors give insights into the single-particle structure approaching the shape transition. 
In \nuc{94}{Sr} the selectivity of the $(d,t)$ reaction allowed for the identification of excited $0^+$ states. The spectroscopic factors are interpreted as resulting from the mixing of two distinct structures and show that there is strong mixing at $N=56$.

\section{Experimental Details} 

\begin{figure}[h]
 \includegraphics[width=\columnwidth]{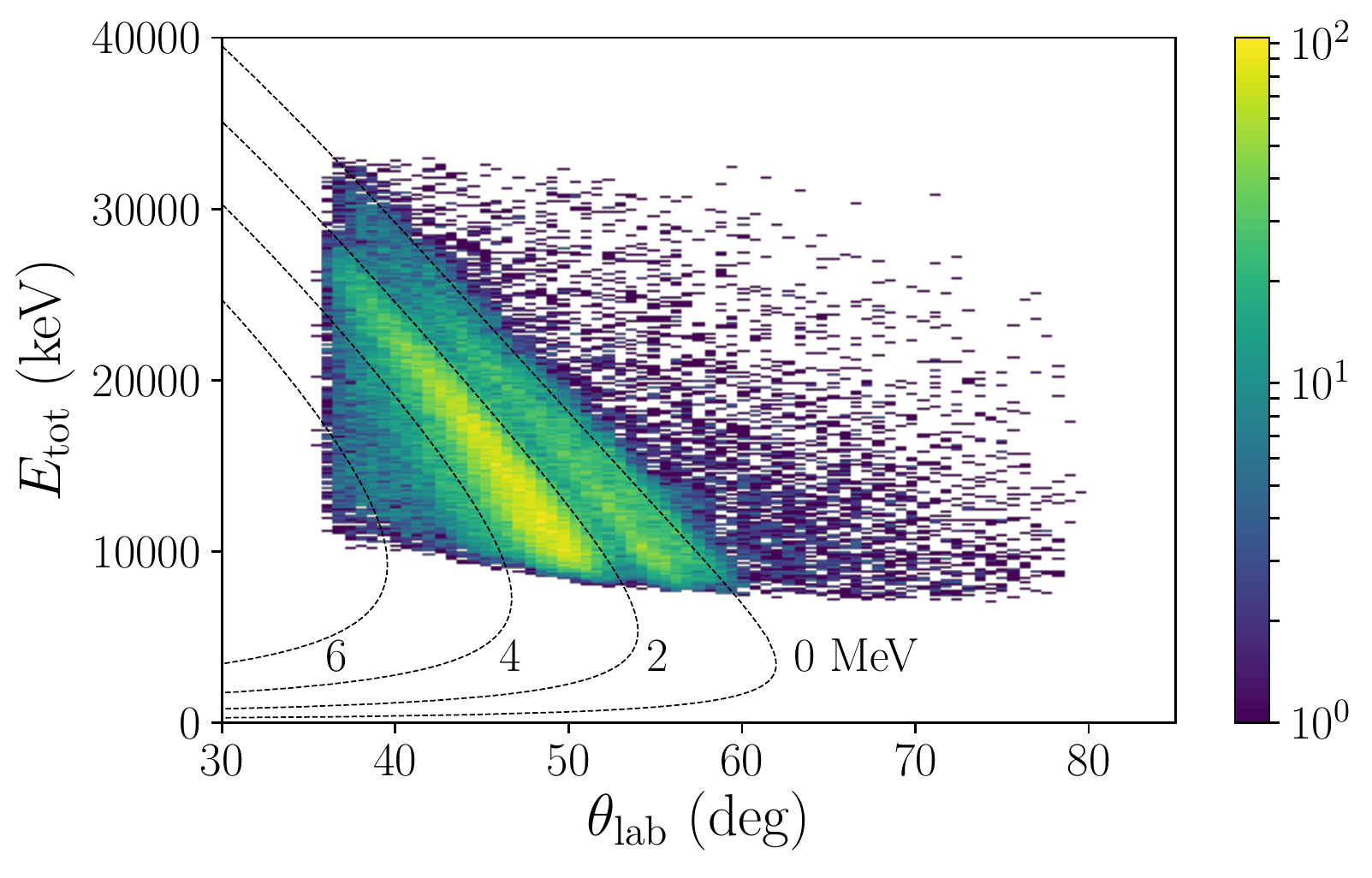}
 \caption{\label{fig:kine} Kinematics plot for the \dtsr{95} transfer reaction (only identified tritons are shown). The transfer to states at excitation energies at 0, 2, 4, 6 MeV are denoted by the dashed lines.}
\end{figure}

The present \dtsr{94,95,96} experiments were carried out at the TRIUMF-ISAC II rare isotope facility~\cite{ball11}. 
The neutron-rich Sr isotopes were produced when a 480 MeV proton beam (with 10 $\mu$A current) bombarded a thick uranium carbide (UC$_\text{x}$) target. 
After element selective two-step resonant laser excitation and ionization of Sr with the TRIUMF laser ion source~\cite{lassen17} the extracted beam was mass separated, charge bred, and accelerated to 5.5~$A$MeV using the ISAC I and II heavy ion accelerators. 
Details of the beam production can be found in~\cite{cruz18,cruz19}.
The post-accelerated beams were impinging upon $\approx0.5$~mg/cm$^2$, CD$_2$ deuterated polyethylen targets ($>90$~\% deuterium/hydrogen ratio) at the center of the SHARC/TIGRESS setup.
The light charged particles from elastic scattering, fusion, and transfer reactions were detected and identified by using the highly segmented silicon detector array (SHARC)~\cite{diget11}. 
SHARC provides good angular coverage ($35^\circ<\theta_\text{lab}<80^\circ$ downstream of the target) and resolution ($\Delta \theta_\text{lab} \approx 1 ^\circ$), ideal for measurements of angular distributions of light ion induced reactions.
A detailed description of the configuration and performance of the array is presented in~\cite{cruz19}. For the present work, only the laboratory forward detectors were used as the $(d,t)$ reaction kinematics restricts the scattering angles for the tritons to $\lesssim50-60^\circ$. The downstream section of SHARC is composed of silicon $\Delta$E - E detectors (thickness $\approx$ 150 $\mu m$ and $\approx$ 1500 $\mu m$) covering the angular range 35$^{\circ} < \theta_{lab} <$ 80$^{\circ}$ in the laboratory system. The setup allowed for the identification of tritons with kinetic energy above about 8~MeV. The particle identification plot is presented in Fig.~1 of~\cite{cruz19}. 
The kinematics plot for the \dtsr{95} reaction is shown in Fig.~\ref{fig:kine}.

The $\gamma$ rays emitted from excited states in the Sr residue nuclei were detected by TIGRESS (TRIUMF-ISAC Gamma-Ray Escape Suppressed Spectrometer)~\cite{hackman14} consisting of twelve segmented clover detectors surrounding the SHARC array. 
Eight (four) clover detectors were placed at 90$^{\circ}$ (135$^{\circ}$) with respect to the beam direction.
The distance from the target to the front face of the detectors in the high-efficiency configuration is 11.0~cm resulting in a full-energy peak efficiency of $\approx$ 10~\% at a $\gamma$-ray energy of 1~MeV.

\section{Data Analysis and Results} \label{sec:ana}
The data collected with SHARC were calibrated using a standard triple-$\alpha$ source ($^{239}$Pu-$^{241}$Am-$^{244}$Cm) as well as light ions from elastic scattering, \ddsr{} and \ppsr{}. The excitation energies of final states in the Sr transfer residues were derived from the measured energies and angles of identified tritons. The in-beam excitation energy resolution was limited to $\sigma \approx 270$~keV due to beam energy loss in the thick target. This resolution was determined from isolated (or $\gamma$-ray gated) peaks in the excitation energy spectrum (see Fig.~\ref{fig:excfit95}). It is almost independent of the scattering angle for the range covered in this experiment. 
The energy and efficiency calibrations of the TIGRESS Ge detectors were carried out using standard radioactive $^{152}$Eu and $^{60}$Co sources, placed at the target position.
Level schemes were constructed using triton-$\gamma$ (and triton-$\gamma$-$\gamma$ where possible) coincidences. Gating on the excitation energy allows one to associate $\gamma$ rays with excited states, while the reverse gating gives information on whether the state was directly populated or fed from above.

The differential cross sections were analyzed with finite-range distorted wave Born approximation (DWBA) calculations using the FRESCO code~\cite{thompson88}.
It is assumed that in the present \dtsr{94,95,96} experiments the final states are populated in a single-step process in which a
valence neutron is removed from the \nuc{94,95,96}{Sr} ground states.
In order to determine the absolute cross section the luminosity of each experiment was determined by fitting the elastic scattering data as shown in Fig.~2 of Ref.~\cite{cruz19}. The optical model parameters for the deuteron-nucleus potentials were also determined for the \ddsr{94,95,96} data. They are given in Table~II of Ref.~\cite{cruz19}. For the triton-nucleus potential a global parametrization~\cite{li07} was employed. The $(d,t)$ overlap function was taken from Green's function Monte-Carlo calculations which give a $(d,t)$ spectroscopic factor of 1.30~\cite{brida11,*gfmc}. For the Sr nuclei, the bound state wave functions were calculated in a Woods-Saxon potential with $r=1.3$ and $a = 0.66$. The depth of the well was adjusted to match the binding energy of the state. The $1g_{9/2}$, $2d_{5/2}$, $3s_{1/2}$, $1g_{7/2}$, and $2d_{3/2}$ single-particle orbitals were considered in the analysis. Neutron removal from the $1g_{9/2}$ orbital below $N=50$ is, however, unlikely to contribute to the population of low-lying states in the Sr nuclei.
Orbital angular momentum transfers and spectroscopic factors were obtained by comparing the experimental differential cross sections with the reaction model calculations. This analysis carries substantial systematic uncertainties in the reaction modeling, the optical model parameters, and the potential used to calculate the nucleon bound-state wave function.
The normalization of the elastic scattering cross section used to determine the absolute transfer cross sections depends on the optical model parameters chosen for the deuteron-nucleus potentials. When the analysis is restricted to the most forward angles in the laboratory system, the different parametrizations result in a 3~\% systematic uncertainty for the absolute cross sections. The effect of the optical model parameters has been estimated by calculating the spectroscopic factors for different parametrizations taking into account the angular range of covered in the experiment. These vary by 10~\% for $L=0$ and 13~\% for $L=2$, respectively. Lastly, the potential used to calculate the wave functions, and in particular the radius assumed, affect the extracted spectroscopic factors. By varying the parameters of the Woods-Saxon potential within 10~\%, the uncertainty for the spectroscopic factors was estimated as 13~\% for $L=0$ and 22~\% for $L=2$ in the angular range covered by the SHARC detectors. The total systematic uncertainty related to the reaction modeling thus amounts to 17 and 26~\% for $L=0$ and 2, respectively. The spectroscopic factors are quoted below only with the statistical uncertainty. For the absolute comparison with the shell model calculations or other works, the systematic uncertainty has to be considered in addition. 

\subsection{\nuc{93}{Sr}}
\begin{figure}[h]
\includegraphics[width=\columnwidth]{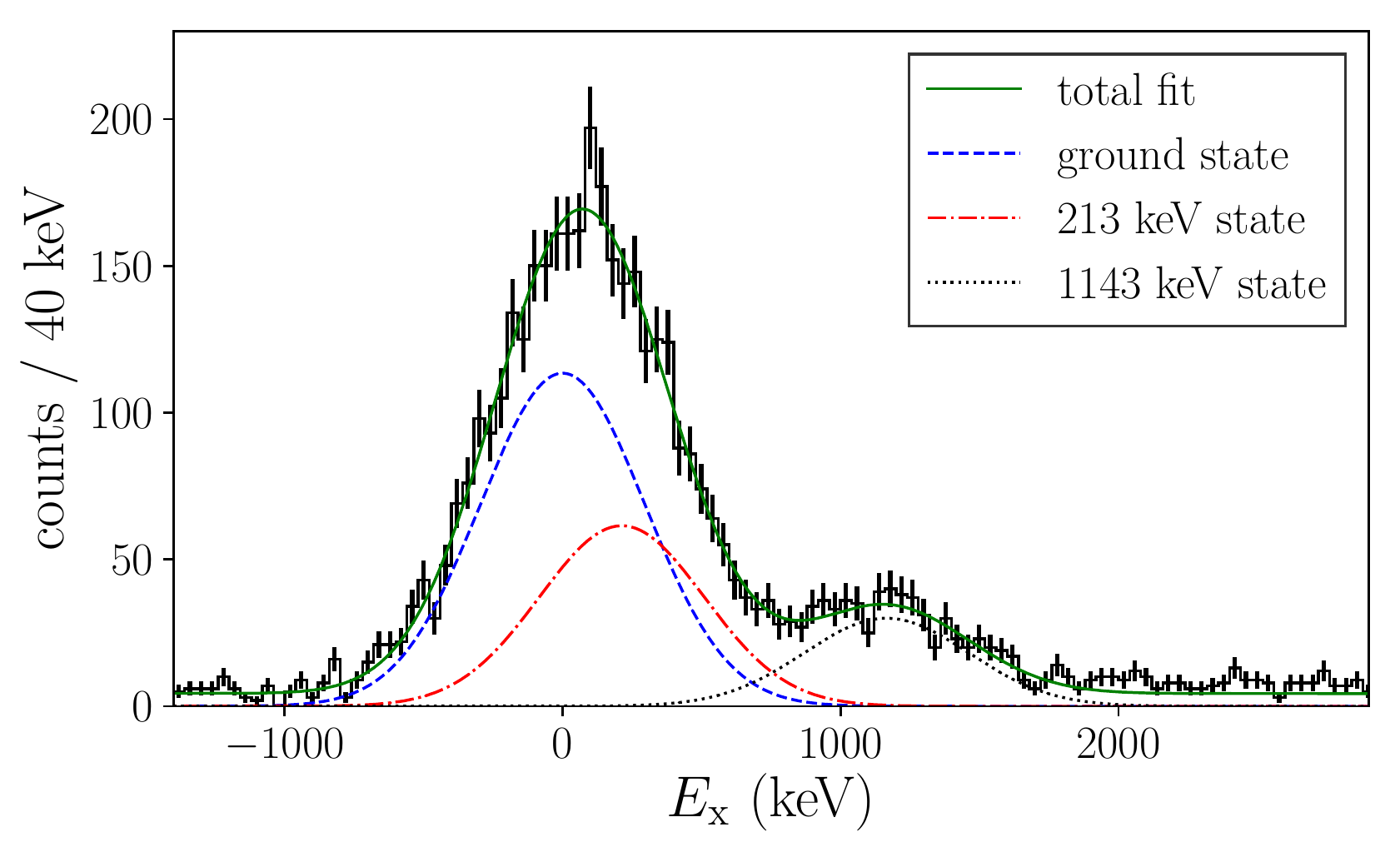}
\caption{The total excitation energy ($E_\text{x}$) spectrum for states in \nuc{93}{Sr} populated in the \dtsr{94} reaction. The excitation energy spectrum was fitted with three components corresponding to the observed states at 0, 213, and 1143~keV. A small contribution from \nuc{94}{Y} and \nuc{93}{Rb} appearing a negative excitation energies in the spectrum was subtracted (see text).}
\label{fig:excfit94}
\end{figure}

\begin{figure}[h]
\includegraphics[width=\columnwidth]{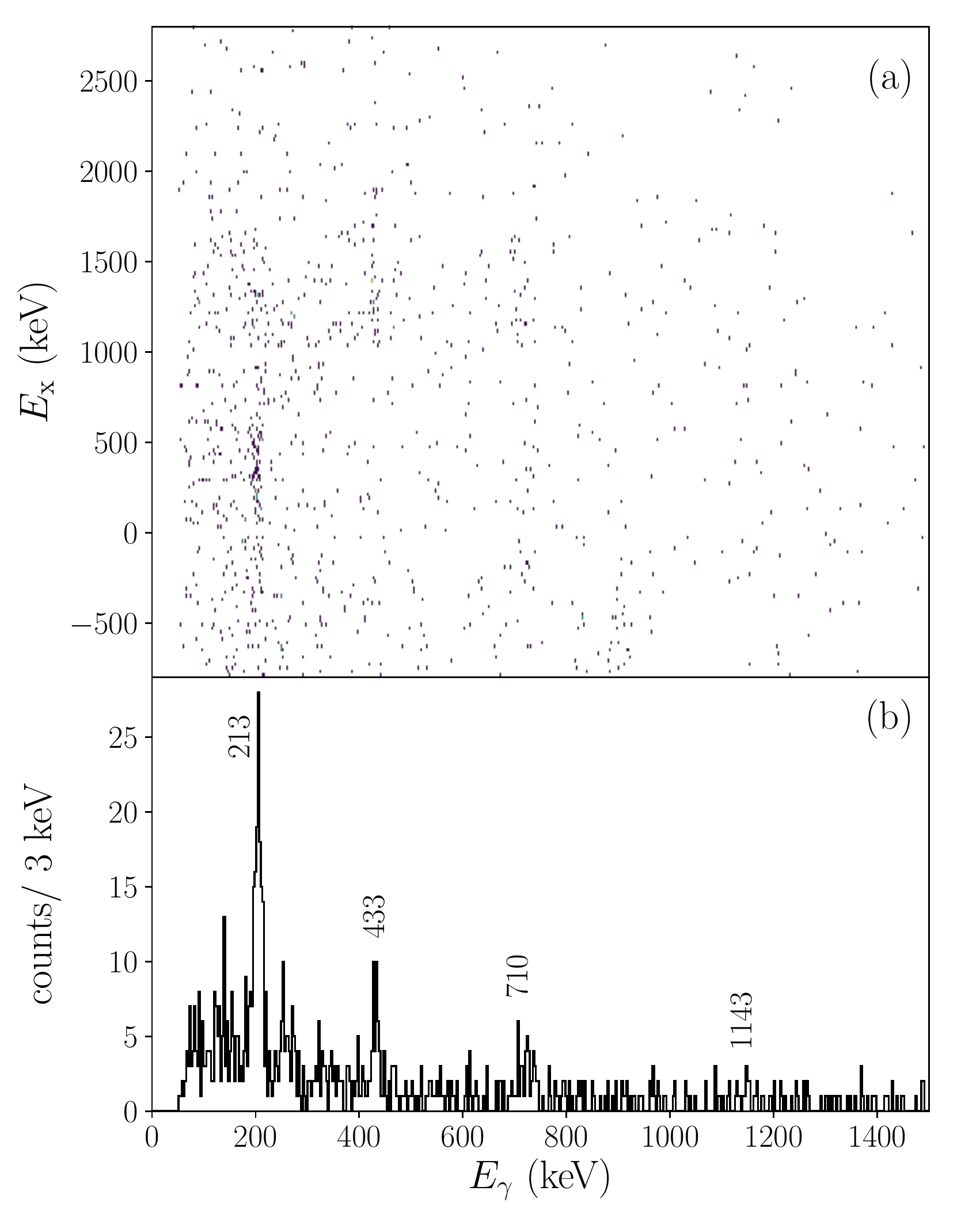} 
\caption{\label{fig:gam94}(a) Excitation energy ($E_\text{x}$) versus $E_{\gamma}$ spectrum obtained from the \dtsr{94} reaction. (b) corresponds to the $\gamma$-ray energy spectrum.}
\end{figure}

\nuc{93}{Sr} was populated via the \dtsr{94} reaction. The excitation energy spectrum is shown in Fig.~\ref{fig:excfit94}.
As evident from the $\gamma$-ray spectra shown in Fig.~\ref{fig:gam94}, the ground and excited states at 213 and 1143~keV have been populated directly in the reaction. The excitation energy spectrum has been fitted with three Gaussian peaks. The separation between the centroids is constrained using the excitation energies and the widths are constrained in accordance with the excitation energy resolution (see Fig.~\ref{fig:excfit95} for the \dtsr{95} reaction). Additionally, a constant background has been added to the fit, representing background events from fusion evaporation reactions.
The excitation energy $E_\text{x}$ - $E_{\gamma}$ and the $\gamma$-ray energy spectra are presented in Fig.~\ref{fig:gam94}.
As shown in Fig.~\ref{fig:gam94} the 433~keV $\gamma$ rays are in coincidence with population of the 1143~keV state. The 433~keV state has not been directly populated in the \dtsr{94} reaction. While the 219~keV transition from this state to the 213~keV state cannot be resolved, the intensity of the 710~keV and 433~keV $\gamma$-ray transitions as function of the excitation energy allow extraction of the direct population of the 433~keV state. A $2\sigma$ upper limit of 4~\% of the population of the 1143~keV state was determined. The 213~keV state is almost exclusively populated directly, a small contribution of indirect feeding from above is observed in the $\gamma$-gated excitation energy spectrum. Note that this state is isomeric with a half-life of $T_{1/2} = 4.3(1)$~ns~\cite{sasanuma04}, therefore the efficiency for detecting this transition in the present experiment is much reduced, similar to the case of the decay of the $0^+_2$ state in \nuc{96}{Sr}~\cite{cruz18}. No evidence for the population of the 986~keV $J^\pi = (9/2^+)$ state was found. In the $\gamma$-ray energy spectrum gated on negative excitation energy a small contribution from \nuc{93}{Rb} was found. 
This contribution has been subtracted by gating on the respective transition energies and subtracting an efficiency-scaled Gaussian from the distribution as shown in Fig.~\ref{fig:excfit94}. This subtraction affects the relative contribution of the ground and 213~keV state to the peak. The relative population is also affected by the energy calibration, but for the spectroscopic factors (see below) only the sum of all contributions was taken into account.
The observed $\gamma$ rays were placed in the level scheme as presented in Fig. \ref{fig:level93}.

\begin{figure}[h]
\includegraphics[width=\columnwidth]{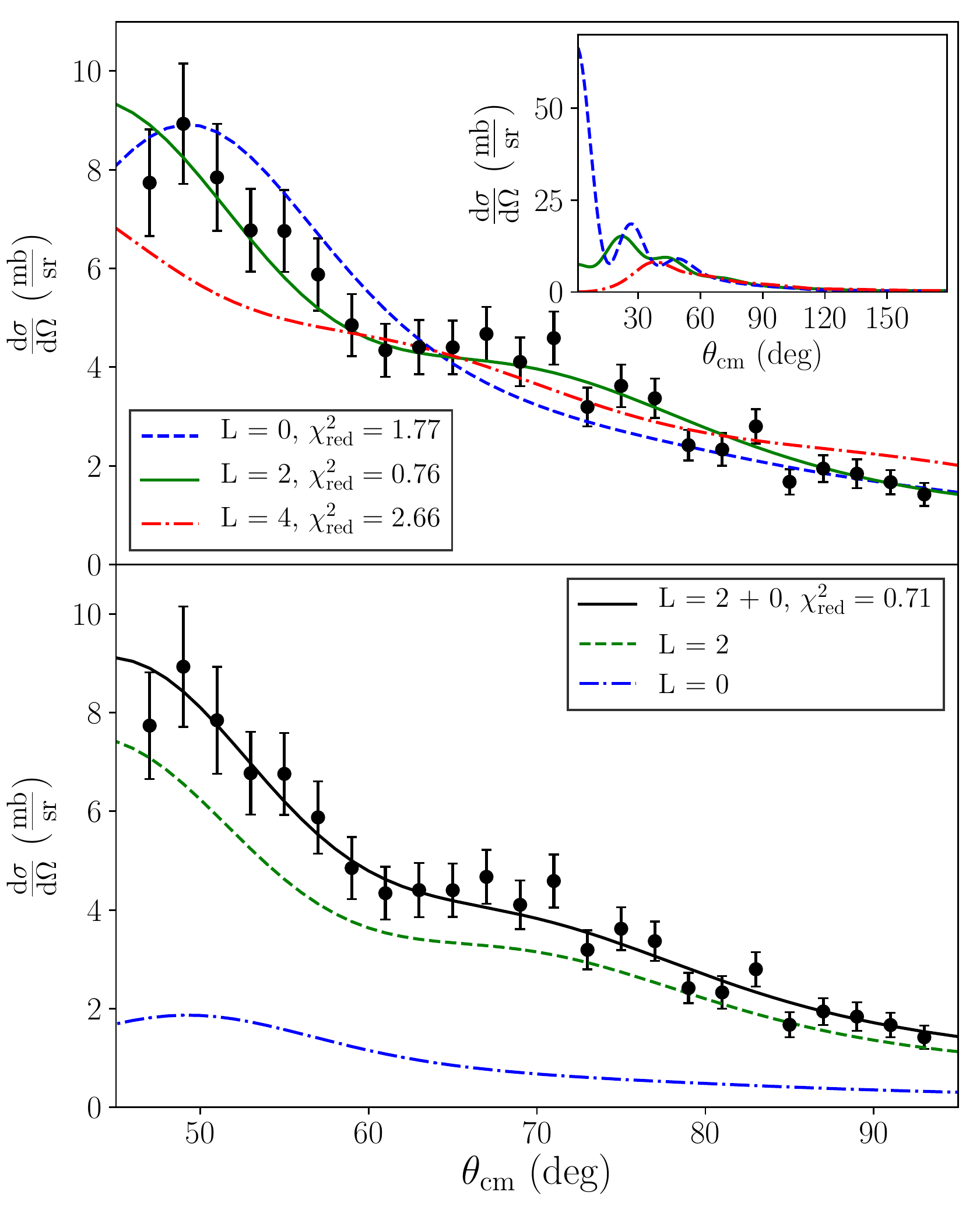}
\caption{\label{fig:angdist94} Differential cross sections for states populated in the \dtsrf{94}{93} reaction. (a) Angular distribution of tritons gated on both the ground and 213~keV states. The data are compared to DWBA reaction model calculations assuming removal of a neutron from the $3s_{1/2}$ orbital with $L=0$ (blue), the $2d_{5/2}$ $L=2$ orbital (green), and the $L=4$ $1g_{9/2}$ orbital (red). The inset shows the full angular range. (b) Same data fitted with a combination of $L=0$ and $L=2$ (solid black line).}
\end{figure}

The limited resolution for the excitation energy and the insufficient statistics for $\gamma$-ray coincidences does not allow extraction of the angular distributions for individual states. Instead the combined angular distribution for the ground and 213~keV states was extracted. These states are rather well separated from the 1143~keV state. The data are compared to DWBA reaction model calculations assuming removal of a neutron from the $3s_{1/2}$, $2d_{5/2}$, and $1g_{9/2}$ orbitals in Fig.~\ref{fig:angdist94} (a).
The data are in good agreement with calculations assuming angular momentum transfer $L = 2$, with exclusion of pure $L=0$ or $L=4$ distributions with 98.71 and 99.99~\% probability, respectively.
Using a two component fit the angular distribution shown in Fig.~\ref{fig:angdist94} does not allow for an additional $L=4$ component, a $J^\pi = 9/2^+$ assignment for the 213~keV state can thus be excluded. Furthermore, the cross section for the population of the 213~keV state is rather high and population of hole states in the deeply bound $1g_{9/2}$ orbital are not expected at low excitation energy. Fitting the angular distribution with a combination of $L=2$ and $L=0$ slightly improves the reduced $\chi^2$ value. The combined fit together with the individual components is shown in Fig.~\ref{fig:angdist94} (b). The spin and parity of the 213~keV state is thus assigned as $(1/2)^+$, based on the present work only, while $(3/2,5/2)^+$ can not be excluded. The spectroscopic factors extracted from the combined fit in Fig.~\ref{fig:angdist94} (b) amount to $C^2S = 3.37(67)$ for the $L=2$ ground state and $0.44(34)$ for $L=0$ for the 213~keV state, where the uncertainties account for the statistical uncertainty only. The systematic uncertainty related to the contamination discussed above amounts to 10~\%. The fit with only the $L=2$ component results in a summed spectroscopic factor of $C^2S = 4.24(11)$. The spectroscopic factors are summarize in Table~\ref{tab:specfac}. Regarding the 1143~keV state, statistics are not sufficient to conclude on the spin and parity from the the angular distribution. Assuming $L=2$ the spectroscopic factor for this state amounts to $C^2S = 0.65(15)$.

\subsection{\nuc{94}{Sr}}
\begin{figure}
\includegraphics[width=\columnwidth]{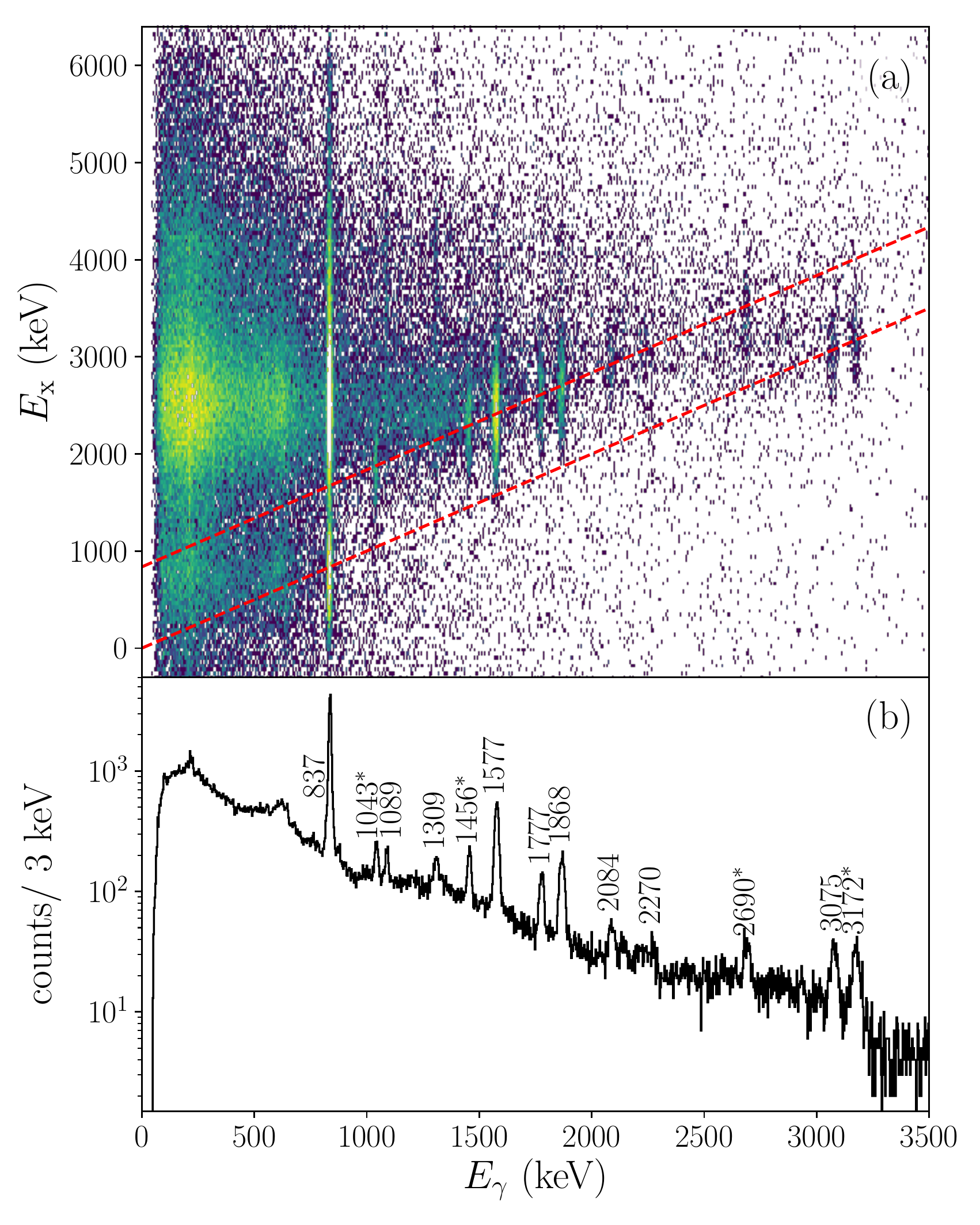}
\caption{\label{fig:gam95} (a) Excitation energy ($E_\text{x}$) versus $E_{\gamma}$ spectrum for \nuc{94}{Sr} populated via the \dtsr{95} reaction. The dashed lines indicate the location of direct ground state decays and decays to the first excited $2^+$ state at 837~keV. (b) corresponds to the $\gamma$-ray spectrum. The newly observed transitions are marked with asterisks.}
\end{figure}

Low-spin states in \nuc{94}{Sr} were populated by the \dtsr{95} reaction. In this case much higher statistics were obtained such that the level scheme was constructed using the triton-$\gamma$ coincidences shown in Fig.~\ref{fig:gam95} as well as $\gamma$-$\gamma$ coincidences.
The excitation energy ($E_\text{x}$) versus $E_{\gamma}$ spectrum in Fig.\ref{fig:gam95} (a) shows the direct population of several states. The transitions at 1043, 1456, 2690, 3075, and 3172~keV are observed for the first time. It is evident from Fig.~\ref{fig:gam95} (a) that the 837, 3075, and 3172~keV transitions are direct ground state decays. All other transitions feed the first excited $2^+$ state at 837~keV. This was further verified using $\gamma-\gamma$ coincidences.

\subsubsection{The $0^+$ states}
\begin{figure}[h]
\includegraphics[width=\columnwidth]{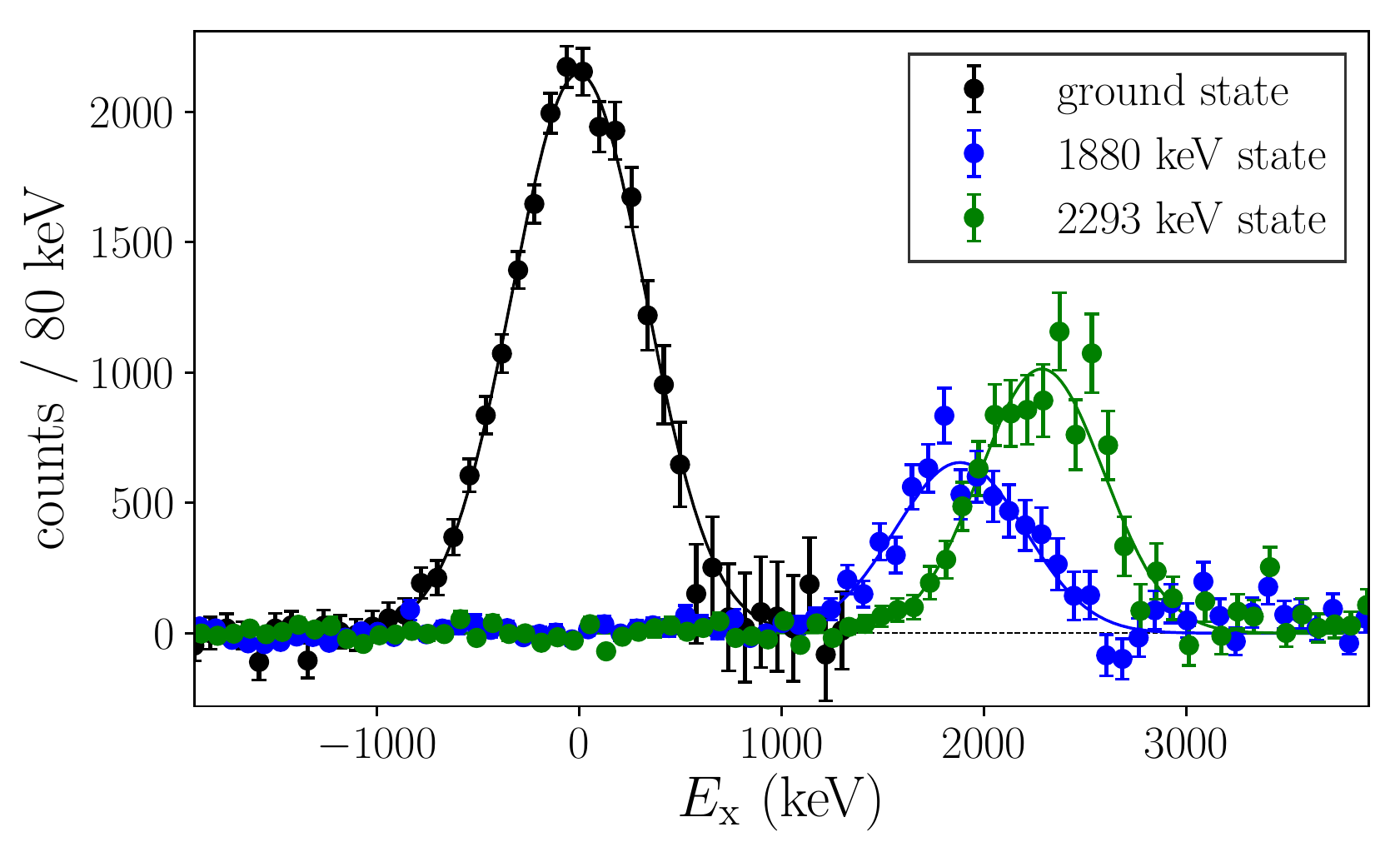}
\caption{\label{fig:excfit95}Exclusive excitation energy ($E_\text{x}$) spectra for the the \zps\ populated in the \dtsrf{95}{94} reaction. The ground state was extracted from the total spectrum by removing contributions from the 837~keV state using a $\gamma$-ray gate. A linear background was estimated using the negative excitation energy region and subtracted from the spectrum. For the excited states, a gate was placed on the corresponding $0^+_x \rightarrow 2^+_1$ transitions and the spectrum was scaled by the respective $\gamma$-ray detection efficiency. The Gaussian fits were used to determine the excitation energy resolution.}
\end{figure}

\begin{figure}[h]
\includegraphics[width=\columnwidth]{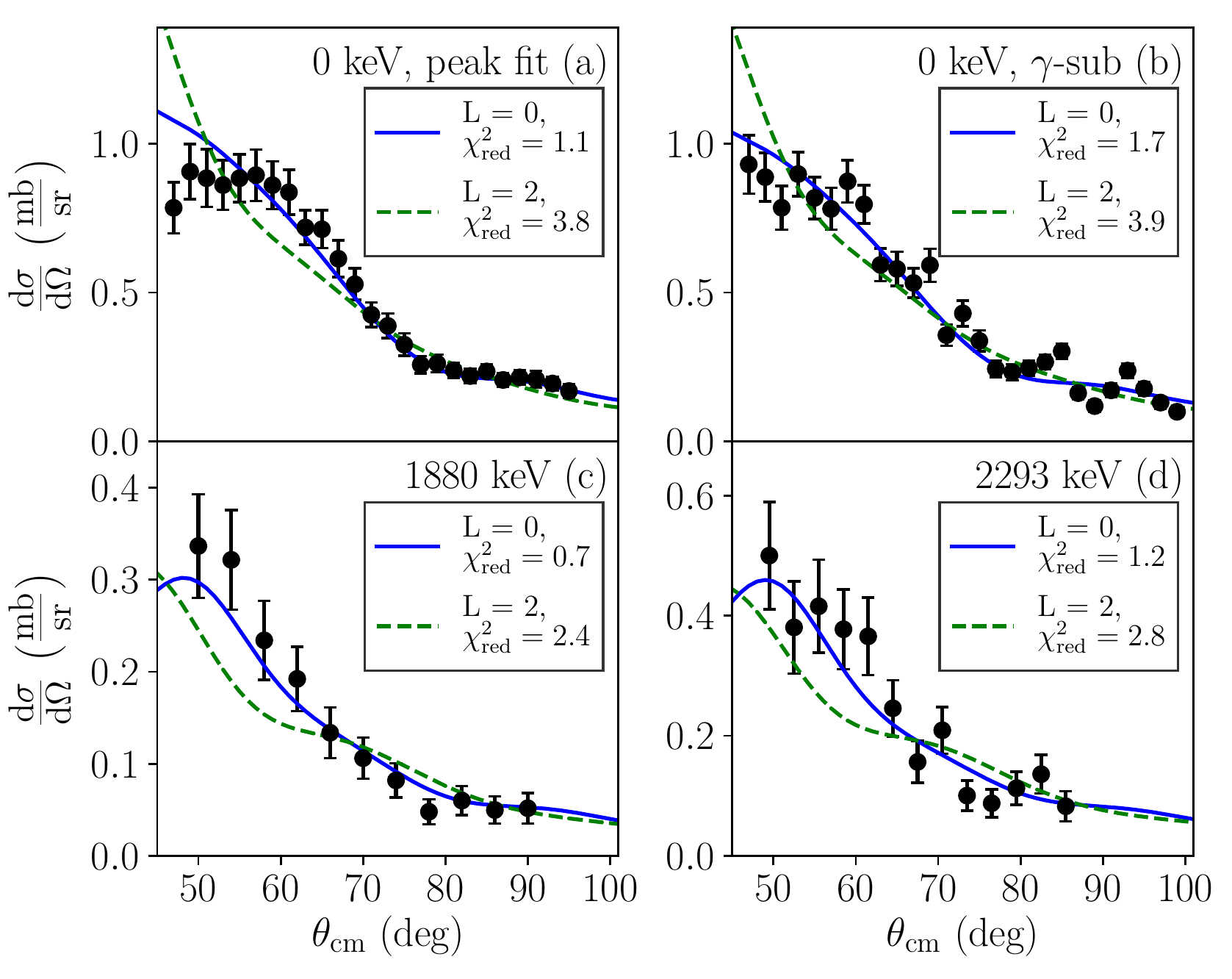}
\caption{\label{fig:angdist95_0} Angular distribution of tritons for $L=0$ transfer reactions to final states in \nuc{94}{Sr}. (a) and (b) show the results of the two methods to extract the ground state distribution. The newly discovered $0^+$ states are shown in (c) and (d). The $L=0$ and $L=2$ DWBA calculations are fitted to the data, and are shown by the blue (solid) and green (dashed) lines and labeled with their respective reduced $\chi^2$ values.}
\end{figure}

The ground state of \nuc{94}{Sr} was strongly populated in the present experiment. The angular distribution for the reactions to the ground state was extracted using two methods. Firstly, a constrained Gaussian peak fit was employed to fit both the ground state and the 837~keV state for each angular bin. The width of the Gaussians has been fixed to $\sigma = 270$~keV. In the second method, the distribution gated on the 837~keV transition was subtracted from the combined angular distribution, after scaling for efficiency. An additional gate was placed on excitation energies $E_\text{x}<1.4$~MeV to exclude states that decay directly to the ground state. This excitation energy spectrum is shown in Fig.~\ref{fig:excfit95}.
The angular distribution for the reactions to the ground state extracted using both methods is shown in Fig.~\ref{fig:angdist95_0} (a) and (b).
The differential cross sections are fitted with DWBA calculations described in Section~\ref{sec:ana}. The data extracted both ways clearly favors a calculation with angular momentum transfer $L=0$. This corresponds to the removal of a $3s_{1/2}$ neutron from the $J^\pi=1/2^+$ ground state of \nuc{95}{Sr}. The spectroscopic factors amount to $C^2S = 0.336(7)$ for the peak-fit method, and $0.314(7)$ for the subtraction method. This value for the spectroscopic factor agrees very well with the one obtained for the \dpsrf{94}{95} reaction measured within the same campaign ($C^2S = 0.41(9)$~\cite{cruz19}).

The newly observed 1043 and 1456~keV transitions were found to be in coincidence with the 837~keV $2^+_1 \rightarrow 0^+_1$ transition. The excitation energy spectra gated on these two transitions are shown in Fig.~\ref{fig:excfit95}. It is clear that they are arising from the population of states at excitations energies of 1879.7(6) and 2292.8(6)~keV. Direct decays to the ground state have not been observed. The upper limit for the non-observation of 1880~keV (2993~keV) $\gamma$-rays relative to the 1043~keV (1456~keV) transition amount to 2~\% (1~\%). The extraction of an upper limit for the ground state decay of the 1880~keV transition is complicated by the nearby 1867~keV transition.
The angular distributions of tritons following the population of the 1880 and 2293~keV states are shown in Fig.~\ref{fig:angdist95_0} (c) and (d). In both cases the agreement with the $L=0$ calculations is significantly better than for $L=2$. This suggests that the states have spin and parity of $0^+$ or $1^+$. The population of a $1^+$ state by $L=0$ transfer would require seniority $\nu\not= 0$ components in the ground state wave function of \nuc{95}{Sr} which results in a small spectroscopic factor. As discussed below such states are expected in the shell model at higher excitation energies.
Together with the decay branching limit to the ground state, $J^\pi = 0^+$ is therefore assigned to both states. 
Spectroscopic factors extracted from the DWBA analysis amount to $0.067(4)$ and $0.105(6)$ for the 1880 and 2293~keV states, respectively.

\subsubsection{$2^+$ states}
\begin{figure}[h]
\includegraphics[width=\columnwidth]{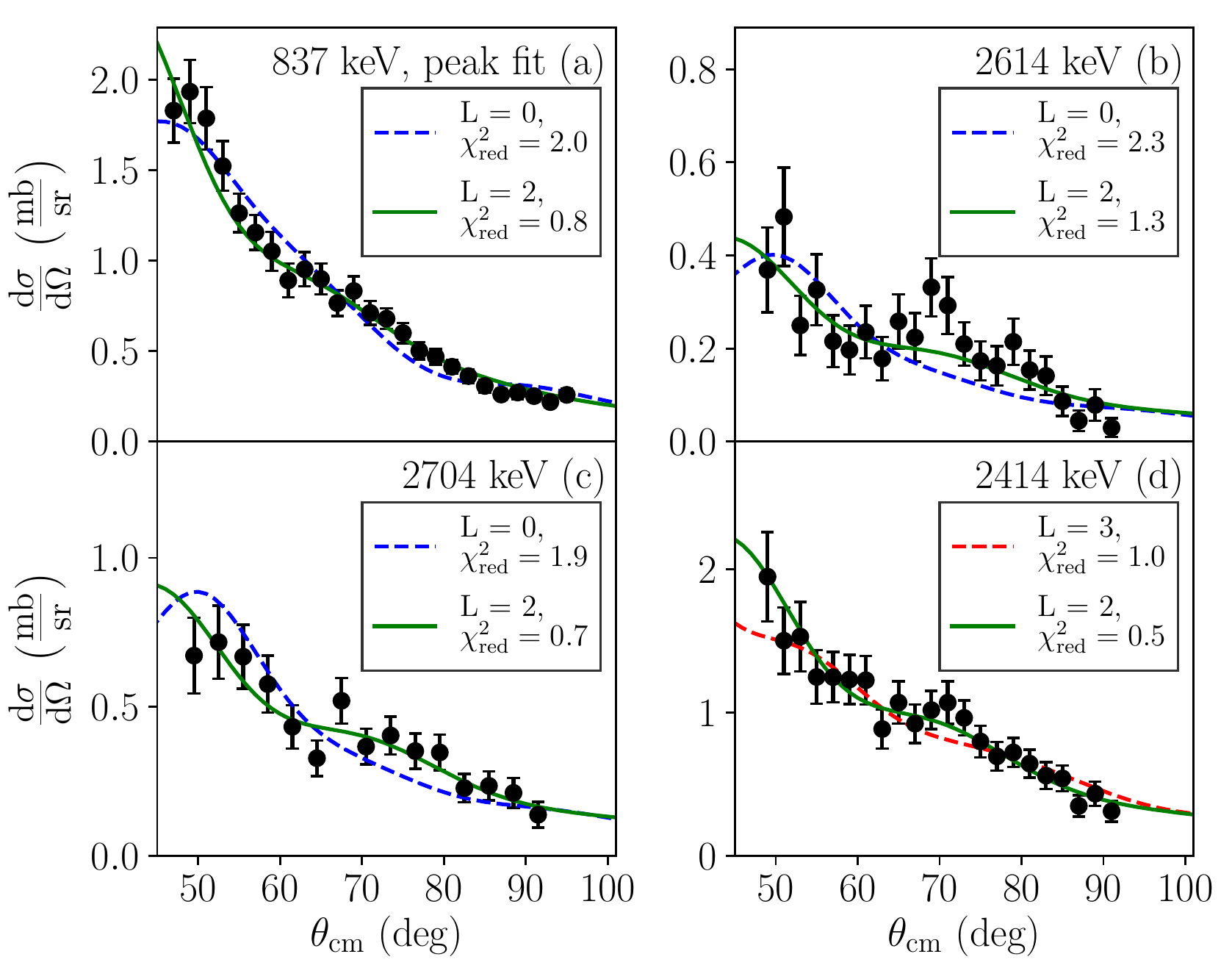}
\caption{\label{fig:angdist95_2} Angular distribution of tritons for $L=2$ transfer reactions to final states in \nuc{94}{Sr}. (a) shows the 837~keV $2^+$ state angular distribution extracted with the peak fit method. (b) and (c) show states which were previously discussed as candidates for $2^+$ states. (d) The 2414~keV state was previously assigned tentatively as $3^-$ which is in clear contradiction to the present result with is consistent only with $L=2$. The $L=0$, $L=2$, and $L=3$ DWBA calculations fitted to the data are shown by the blue, green, and red lines and labeled with their respective reduced $\chi^2$ values.}
\end{figure}

The $2^+_1$ state has been directly and indirectly populated in the present experiment, and thanks to the separation in the excitation energy spectrum, the angular distribution can be obtained from a $\gamma$-gate or from a combined fit of the ground state and the 837~keV state similar to the case of the ground state angular distribution. The latter is shown in Fig.~\ref{fig:angdist95_2} (a).
The spectroscopic factor extracted from the fit with the $L=2$ DWBA calculation amounts to $C^2S = 0.725(15)$.

The two states at 2614 and 2704~keV have sufficient statistics to analyze the angular distributions. They are shown in Fig.~\ref{fig:angdist95_2} (b) and (c), respectively. The spectroscopic factors amount to 0.21(1) and 0.45(2), respectively.
The states at 2270, 2921, and 3527~keV are also in agreement with $L=2$, although the angular distributions are less clear. Populating these states via the removal of a $d_{5/2}$ neutron from the $1/2^+$ ground state of \nuc{95}{Sr} restricts the spin and parity to $(2,3)^+$. 
This is consistent with previous assignments~\cite{jung80}. The 2270~keV state directly decays to the ground state, therefore $J^\pi = 2^+$ is preferred. 

\subsubsection{The 2414~keV state}
A state a 2414~keV has been previously assigned as $J^\pi = 3^-$ based on its strong feeding in the $\beta$ decay of \nuc{94}{Rb} and the measured angular correlation~\cite{jung80}. In the present experiment the 2414~keV state has also been strongly populated. A gate on the 1577~keV transition confirms that the state is almost exclusively ($>95$~\%) directly populated in the $(d,t)$ reaction. Population of negative parity states is not expected as all available neutron orbitals are of positive parity. By removing a neutron from the naively fully occupied $\nu 2d_{5/2}$ orbital the coupling with the neutron in the $3s_{1/2}$ orbital results in a $3^+$ state. The differential cross section for the transfer reaction to the 2414~keV state is shown in Fig.~
\ref{fig:angdist95_2} (d). It is compared to a calculation assuming removal of a $2d_{5/2}$ neutron with $L=2$ and also the hypothetical removal of a $1f_{5/2}$ neutron giving rise to the negative parity. While this is still consistent with the data, the much better fit of the $L=2$ calculation together with the strong population supports the assignment of $3^+$ for this state. The spectroscopic factor amounts to $C^2S=0.99(3)$.

\subsubsection{$1^+$ states}
\begin{figure}[h]
\includegraphics[width=\columnwidth]{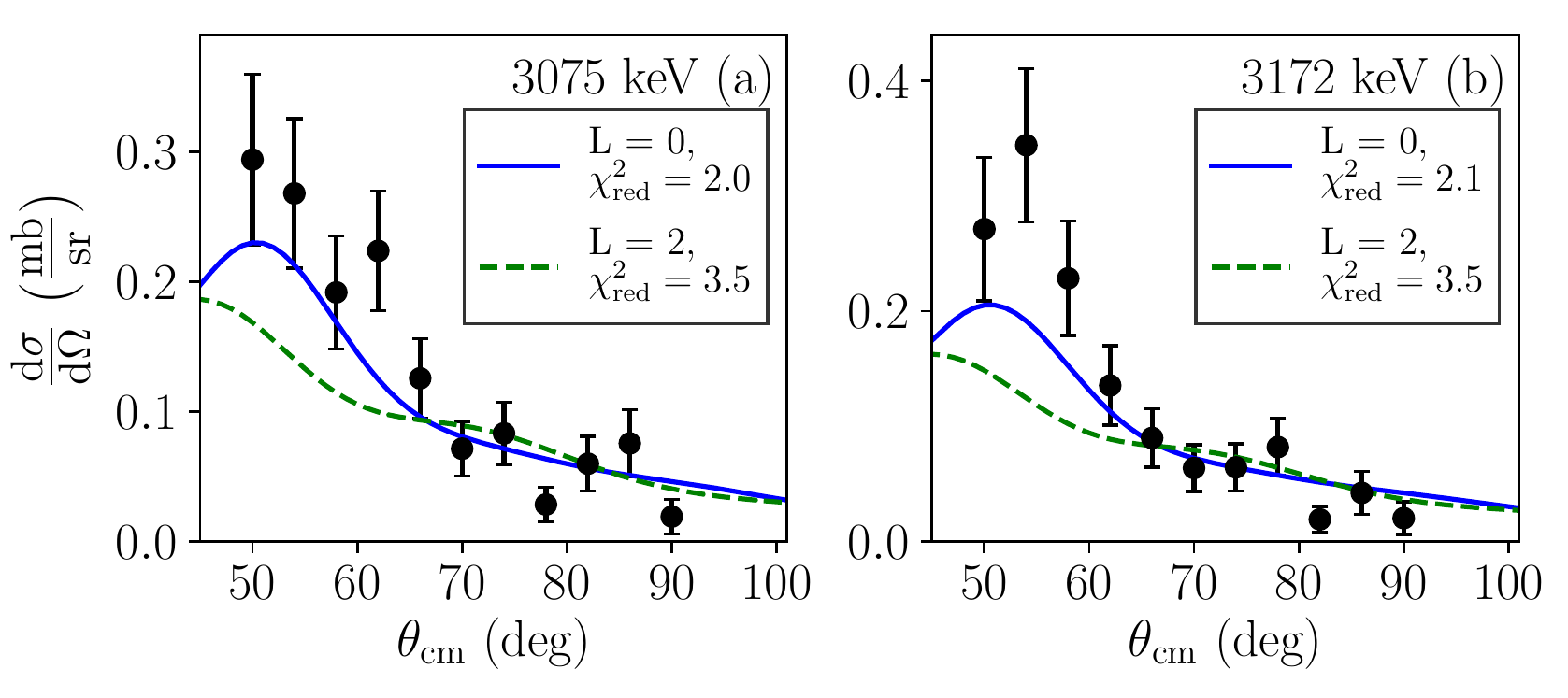}
\caption{\label{fig:angdist95_1} Angular distribution of tritons for the $J^\pi = 1^+$ candidates in \nuc{94}{Sr}. The $L=0$ and $L=2$ DWBA calculations are fitted to the data, and and shown by the blue (solid) and green (dashed) lines and labeled with their respective reduced $\chi^2$ values. Both states clearly agree better with $L=0$.}
\end{figure}

Two states at 3075 and 3172~keV have been observed for the first time in the present work. The former is close to a a previously observed state at 3078~keV~\cite{jungphd}, however, the main branch of the decay of the 3078~keV state is via a 1158~keV transition to the $(3^-)$ state at 1926~keV. The two states observed in the present work decay only directly to the ground state. This indicates that their spins and parities are $(1,2)^+$. Comparison of the differential cross sections with the reaction model calculations shown in Fig.~\ref{fig:angdist95_1} suggests $L=0$ transfer, i.e. removal of a neutron from the $s_{1/2}$ orbital.
Since the ground state wave function of \nuc{95}{Sr} contains a significant contribution of components that have two neutrons in the $3s_{1/2}$ orbital~\cite{cruz19}, such $J^\pi = 1^+$ states can be populated by the reaction. The spectroscopic factors resulting from the fits with $L=0$ shown in Fig.~\ref{fig:angdist95_1} amount to $0.058(5)$ and $0.053(5)$.

\subsubsection{Other states}
While many states in \nuc{94}{Sr} are known, very few have definite spin and parity assignments. Population via the $\beta$ decay of the $3^{(-)}$ ground state of \nuc{94}{Rb} limits the spin to $J=(2,3,4)$ for several states~\cite{jung80}. The $4^+_1$ state at 2146~keV has been only weakly populated directly in the present work. The transition at 2141~keV likely corresponds to the decay of the 2981~keV state to the first $2^+_1$ state.
The population of the $(3^-)$ state at 1926~keV is indirect from unresolved states around 3.5~MeV. A summary of all observed transitions and their placement in the level scheme is presented in Table~\ref{tab:gammatrans}.
\begin{table}[h]         
  \begin{center}       
    \caption{\label{tab:gammatrans} Summary of observed levels, transitions, and relative intensities for the \dtsrf{95}{94}. The transition and level energies with their uncertainties are obtained from the present work, they agree within the uncertainties with the results obtained in $\beta$ decay~\cite{jungphd}. The spin and parity assignments are from the present work.} 
    \begin{threeparttable}
      \begin{tabular*}{\columnwidth}{r@{\extracolsep{\fill}}rrclr}
        \hline
        \hline 
        $E_{\gamma}$ (keV) & I (\%) & \multicolumn{3}{c}{assignment}                    & $E_\text{i}$ (keV) \T\B\\
        \hline
        \hline
        836.7(3)   & 57.0 & $2^+_1  $ & $\rightarrow$ & $0^+_1$ & 836.7(5)   \T\\
        1043.0(5)  & 2.4  & $0^+_2  $ & $\rightarrow$ & $2^+_1$ & 1879.7(6)\tnote{a}  \\
        1089.5(6)  & 1.8  & $(3^-_1)$ & $\rightarrow$ & $2^+_1$ & 1926.2(7)  \\
        1308.8(9)  & 2.4  & $4^+_1  $ & $\rightarrow$ & $2^+_1$ & 2145.5(9)  \\
        2270.4(19) & 0.4  & $(2)^+  $ & $\rightarrow$ & $0^+_1$ & 2270.4(19)  \\
        1456.1(5)  & 3.7  & $0^+_3  $ & $\rightarrow$ & $2^+_1$ & 2292.8(6)\tnote{a}  \\
        1577.4(4)  & 15.0 & $3^+_1  $ & $\rightarrow$ & $2^+_1$ & 2414.1(5)  \\
        1777.1(6)  & 3.3  & $(2,3)^+$ & $\rightarrow$ & $2^+_1$ & 2613.8(7)  \\
        1867.4(4)  & 6.7  & $(2)^+  $ & $\rightarrow$ & $2^+_1$ & 2704.1(5)  \\
        2084.3(17) & 0.9  & $(2,3)^+$ & $\rightarrow$ & $2^+_1$ & 2921.0(17)  \\
        2094.1(20) & 0.4  & $(2,3,4)$ & $\rightarrow$ & $2^+_1$ & 2930.8(20)  \\
        2140.9(28) & 0.6  & $(2,3,4)$ & $\rightarrow$ & $2^+_1$ & 2977.6(28)\tnote{b}  \\
        3075.0(21) & 2.5  & $1^+    $ & $\rightarrow$ & $0^+_1$ & 3075.0(21)\tnote{a}  \\
        3171.7(16) & 1.7  & $1^+    $ & $\rightarrow$ & $0^+_1$ & 3171.7(16)\tnote{a}  \\
        2690.3(21) & 1.3  & $(2,3)^+$ & $\rightarrow$ & $2^+_1$ & 3527.0(21)\tnote{a}  \\
        \hline
        \hline        
      \end{tabular*}
      \begin{tablenotes}\footnotesize
        \item[a] observed for the first time in the present work.
        \item[b] likely corresponds to the transition at 2144.2(2) from the 2981.1(5)~keV state observed in the $\beta$ decay of \nuc{94}{Rb}~\cite{jungphd}.
      \end{tablenotes}
    \end{threeparttable}
  \end{center}
\end{table}

\subsection{\nuc{95}{Sr}}
\begin{figure}[h]
\includegraphics[width=\columnwidth]{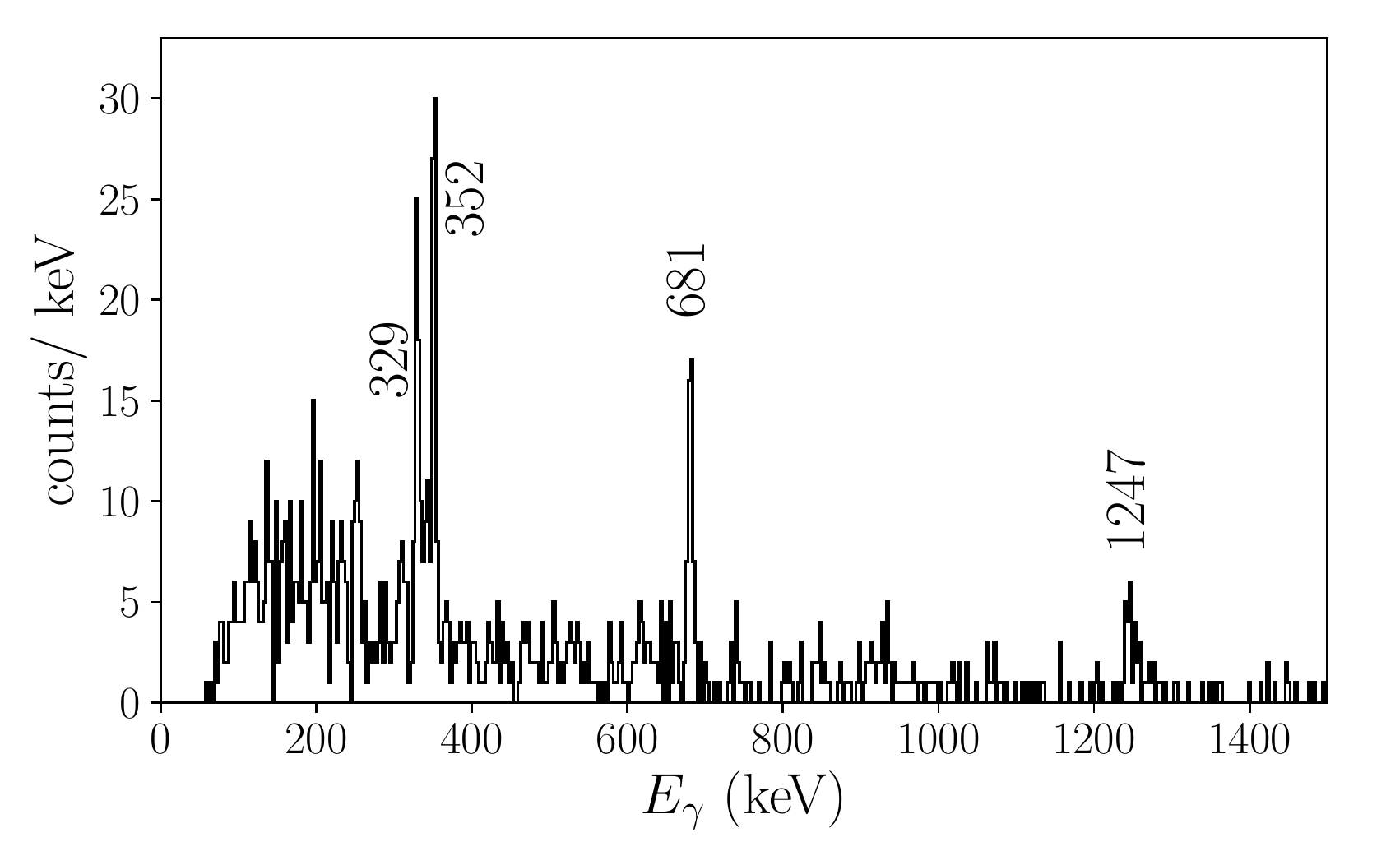}
\caption{\label{fig:gam96} $\gamma$-ray energy spectrum for the \dtsrf{96}{95} reaction. A gate on excitation energies below 1800~keV has been applied.}
\end{figure}

\begin{figure}[h]
\includegraphics[width=\columnwidth]{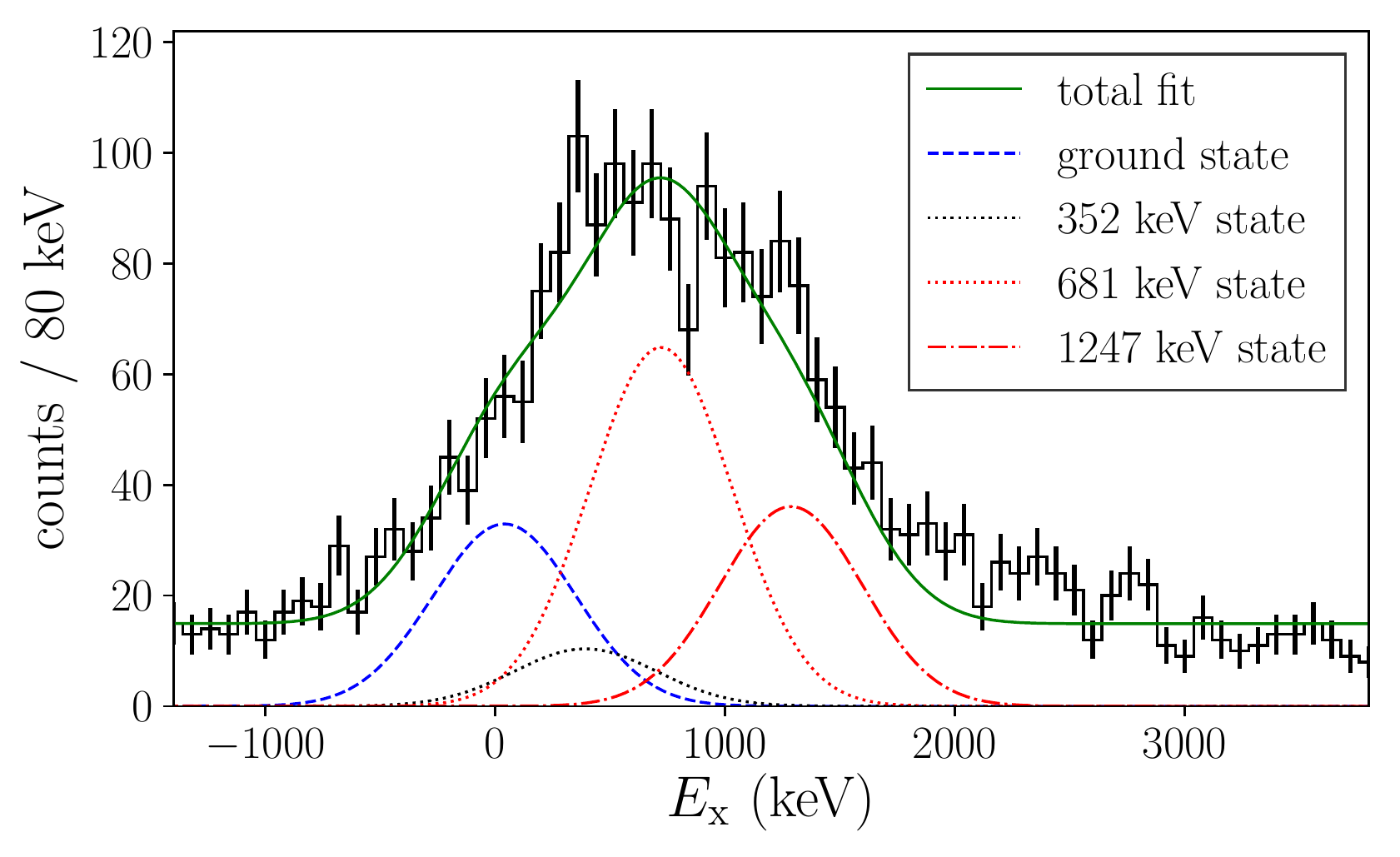}
\caption{\label{fig:excfit96} Excitation energy spectrum for \nuc{95}{Sr}. The total $\gamma$-ray yield has been used to constrain the area of Gaussian peaks and the excitation energy spectrum has been fitted with 4 components and a constant background. The component representing the 352~keV state has been fixed to the upper limit obtained from the $\gamma$-ray analysis.}
\end{figure}

Several states in \nuc{95}{Sr} have been populated in the \dtsrf{96}{95} reaction. The $\gamma$-ray energy spectrum is shown in Fig.~\ref{fig:gam96}.
Four $\gamma$-ray transitions belonging to \nuc{95}{Sr} have been observed. It is evident that the population of the first excited state at 352~keV is mostly indirect via the the 681~keV state which decays by a 329~keV transition to this state. The excitation energy spectrum is shown in Fig.~\ref{fig:excfit96}.
The resolution is not sufficient to separate the close-lying states.
In order to determine the level scheme and the population of the states, the $\gamma$ yield (corrected for efficiency and branching ratios) for bins in excitation energy has been determined. \nuc{95}{Sr} has a transition at 683~keV, belonging to the decay of the 1239~keV state to the isomeric state at 556~keV. Using the known branching ratio of the 681~keV and 329~keV transitions depopulating the 681~keV state~\cite{kratz83}, the yield for the 1239~keV state has been obtained. In contrast to the \dpsrf{94}{95} reaction~\cite{cruz19}, this state is not populated in the present reaction. The direct population strength of the first excited state at 352~keV is consistent with zero. The excess of 352~keV counts in Fig.~\ref{fig:gam96} is likely from states above 1.5~MeV. In order to cross check this analysis and to obtain the ground state population, the excitation energy spectrum has been fitted with four components corresponding to the ground, 352, 681, and 1247~keV states as well as a constant background. The number of counts in the components for the excited states is consistent with the results of a fit to the total $\gamma$-ray energy spectrum shown in Fig.~\ref{fig:gam96}. The result is shown in Fig.~\ref{fig:excfit96}. This allows for the extraction of the population of the ground state  and shows that the data is well described by taking into account only these four states.

\begin{figure}[h]
\includegraphics[width=\columnwidth]{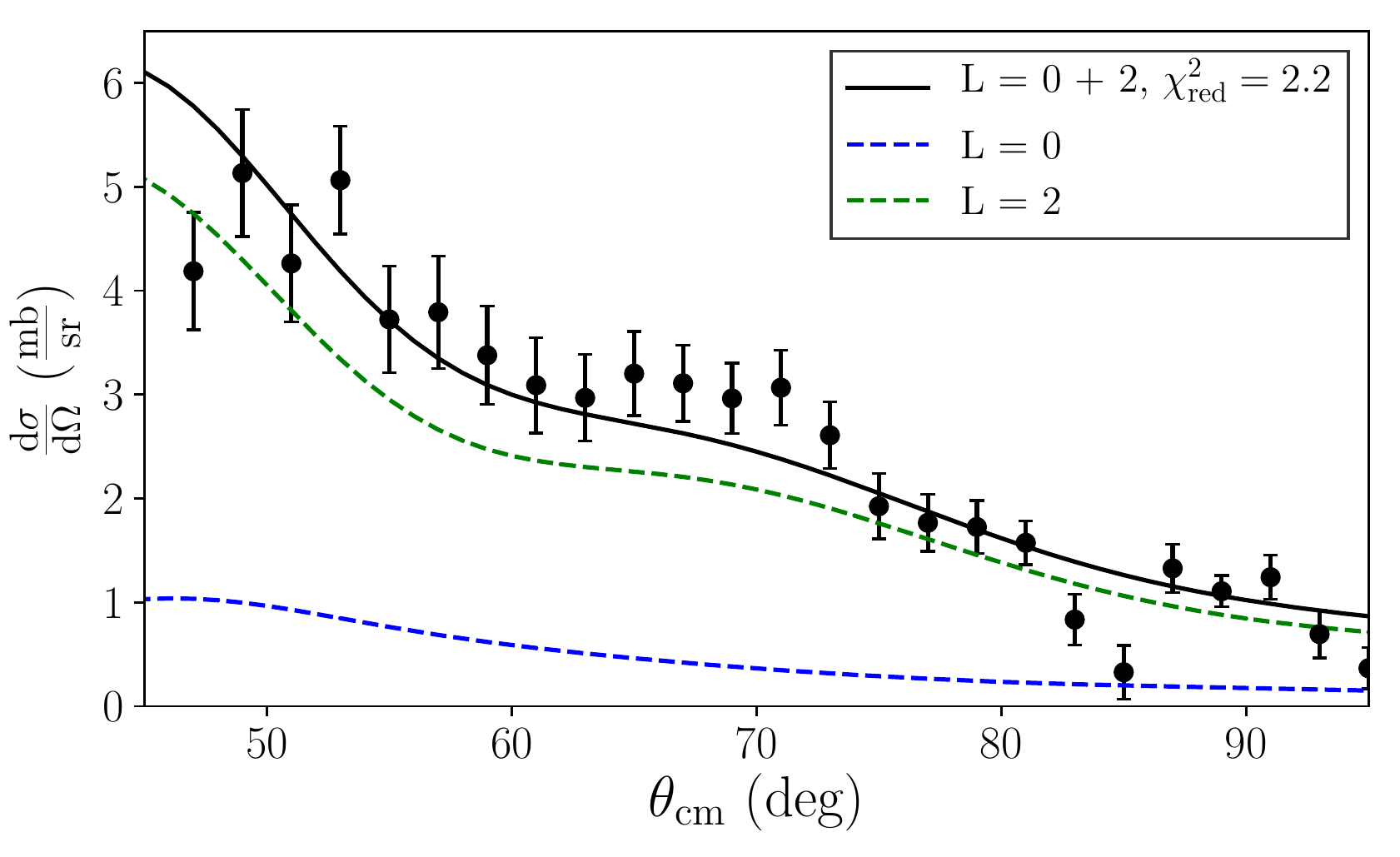}
\caption{\label{fig:angdist96} Angular distribution of tritons gated on both the ground and 618~keV states in \nuc{95}{Sr}. The data are compared to DWBA reaction model calculations assuming removal of a neutron from the $3s_{1/2}$ orbital with $L=0$ for the ground state (blue) and from the $2d_{5/2}$ orbital with $L=2$ for the 681~keV state (green). The combination of $L=0$ and $L=2$ best fitting the data is shown by the solid black line.}
\end{figure}

The angular distribution combined for the ground and 681~keV states has been extracted similarly to the case of the \dtsrf{94}{93} reaction. A gate on excitation energies below 1000~keV has been applied in order to limit the contribution from the 1247~keV state. The spectroscopic factor has been corrected for the fact that the 681~keV state is only partially included in the excitation energy gate. The 352~keV state has been ignored in the angular distribution fit shown in Fig.~\ref{fig:angdist96}. 
The angular distribution for the 1247~keV state has large uncertainties but is best described by $L=0$, which is in agreement with the proposed total angular momentum $(1/2)^+$~\cite{kratz83}. The spectroscopic factors amount to 0.23(15) for the ground state, 2.15(50) for the 681~keV state, and 0.46(15) for the 1247~keV state assuming $J^\pi = 1/2^+$ for the latter.

\section{Discussion} 
\subsection{Structure of \nuc{93}{Sr}}
\begin{figure}[h]
\includegraphics[width=\columnwidth]{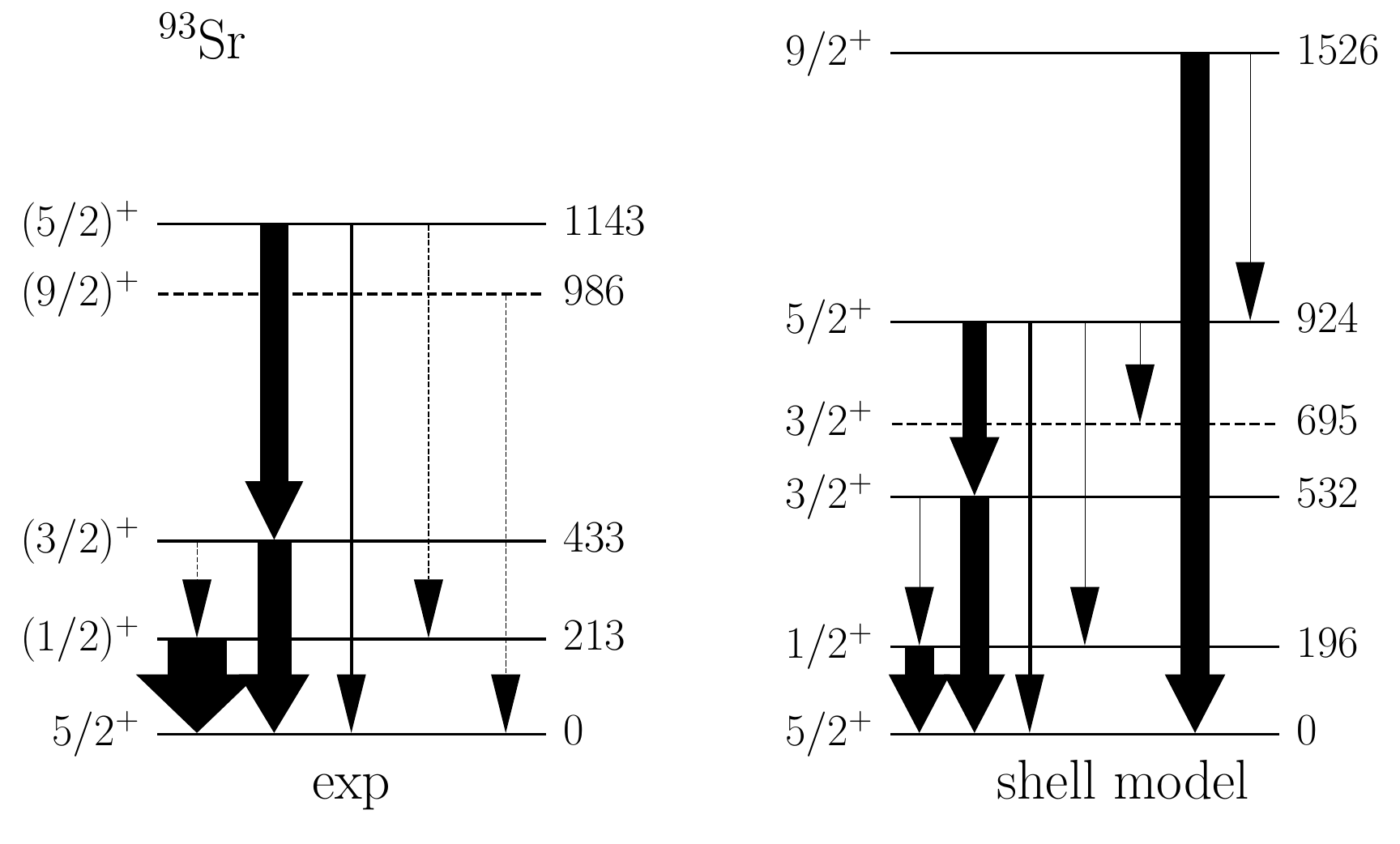}
\caption{\label{fig:level93} Proposed level scheme of \nuc{93}{Sr}. The left side shows the experimental levels with energies in keV and the spin and parity assignments from this work. The widths of the arrows are proportional to the relative intensity of the observed $\gamma$-ray transition. Dashed levels and transitions have not been observed in the present work. The right side shows the result of shell model calculations. Here the widths of the arrows indicate the branching ratios. A $3/2^+$ state at 695~keV cannot be associated with any of the low-lying experimental states.}
\end{figure}

The present data represents the first reaction study of \nuc{93}{Sr}. Excited states have previously been studied only by $\beta$ decay and spin and parity assignments are based on $\log ft$ values. 
The ground state of \nuc{93}{Sr} is known to be $J^\pi=5/2^+$~\cite{buchinger87}. For the 213~keV state both $(9/2^+)$ and $(3/2^+)$ assignments have been proposed, the latter is favored based on the $g$-factor measurement~\cite{sasanuma04}. Other spin and parity assignments for \nuc{93}{Sr} were based on assumptions about the ground state spin of the $\beta$ decay mother \nuc{93}{Rb}~\cite{achterberg74,*brissot75,bischof77}.
The present data limits the possible values for the 213~keV state to $(1/2,3/2,5/2)^+$, with a slight preference for $1/2^+$ based on the measured angular distribution for the ground and 213~keV states. The conversion coefficient for the 213~keV transition has been measured and points to a pure $E2$ transition~\cite{kawade86}, therefore we assign $J^\pi = (1/2)^+$ to the state at 213~keV. The 433~keV state has not been populated directly in the transfer reaction, which is expected for a $3/2^+$ state since the $2d_{3/2}$ orbital is higher in energy and is expected to have a low occupancy in the ground state of \nuc{94}{Sr}. A state at 986~keV, which exclusively decays to the ground state was also not observed in the present experiment. The assignment is consistent with $(9/2^+)$ because direct population in the $(d,t)$ transfer reaction can only populate the $(1g_{9/2})^{-1}$ component of the wave function, and hole states in the deeply bound $1g_{9/2}$ orbital below $N=50$ are unlikely to occur at low excitation energy. This state has been assigned as a member of a rotational band built on the ground state and is interpreted as weak coupling of the $2d_{5/2}$ neutron hole to the $2^+$ state of the \nuc{94}{Sr} core~\cite{hwang03}. The state at 1143~keV decays to the first three states of \nuc{93}{Sr} and direct population in the $(d,t)$ reaction favors positive parity. Therefore, $J^\pi = (5/2)^+$ is tentatively assigned to this state. The proposed spin and parity assignments for \nuc{93}{Sr} are shown in Fig.~\ref{fig:level93}.
The proposed level ordering matches the first few states in the isotone \nuc{95}{Zr}. The first excited state, located at 954~keV, is a $1/2^+$ state and the spectroscopic factor determined from a $(d,t)$ reaction measurement amounts to 0.18~\cite{cohen63}. The spectroscopic factor extracted for the ground state in the $^{96}$Zr$(d,t)^{95}$Zr reaction is 5.75, larger than the present result, $C^2S=3.37(67)$. This suggests that the $N=56$ sub-shell closure is stronger in the Zr isotopic chain than in Sr. The $(1/2)^+$ assignment for the first excited state is in agreement with the $\log ft$ values measured for the decay of \nuc{93}{Rb}~\cite{bischof77} although the present interpretation differs. The $\log ft$ value of 7.9(3) is at the lower limit for known first forbidden unique transitions required for the decay of the $5/2^-$ ground state of \nuc{93}{Rb} to the $(1/2)^+$ state. A total absorption spectrometry measurement suggests an even smaller $\beta$ decay branch to the 213~keV state~\cite{greenwood97} in agreement with the present interpretation. Furthermore, the angular moment sequence of $5/2^+$, $1/2^+$, and $3/2^+$ for the first three states of \nuc{93}{Sr} also matches with the measured conversion coefficients~\cite{kawade86} which indicates $M1$ (and/or $E2$) transitions for the decay of the 433~keV state to the ground and first excited state. As a consequence any state that decays to the 213~keV $(1/2)^+$ state have to have low angular momentum ($1/2^\pm$, $3/2^\pm$, or $5/2^+$). This means that the weak 1566~keV transition observed in the $\beta$ decay of \nuc{93}{Rb} and assigned to the decay of the 1780~keV state is likely misplaced as the latter was observed in the spontaneous fission of \nuc{252}{Cf} and assigned $(11/2^-)$~\cite{hwang03}. 

Fig.~\ref{fig:level93} compares the proposed level scheme to shell model calculations. The shell model calculations have been performed using the GLEK effective interaction~\cite{mach90}. The model space consisted of the neutron $2d_{5/2}$, $3s_{1/2}$, $3d_{3/2}$, and $1g_{7/2}$ and the proton $2p_{3/2}$ and $2p_{1/2}$ orbitals. The single-particle energies of the neutron $2d_{5/2}$, $3s_{1/2}$, $3d_{3/2}$, and $1g_{7/2}$ orbitals have been adjusted, the interaction and model space is the model space (b) used to analyze the $(d,p)$ reactions. This gave the best overall results for \nuc{94-97}{Sr}~\cite{cruz19}.

The results of the shell model calculations are shown on the right side of Fig.~\ref{fig:level93}. The level ordering, energies, and relative transition strengths are in good agreement with the proposed assignment. The $9/2^+$ state belonging to the ground state rotational band is predicted at higher energies. This is also reflected by the excitation energies of the $2^+$ states in \nuc{92,94}{Sr} which are calculated 200-400~keV higher than experimentally observed. The calculated energy of the first $2^+$ is strongly dependent on the proton space used in the calculations~\cite{cruz19}.
The relative \sfacs\ of the first five states are in good agreement with the calculated spectroscopic factors for the \dtsr{94} reaction. The shell model predicts spectroscopic factors of 4.03 and 0.87 for the ground and 213~keV states in good agreement with the experimental result, $3.37(67)$ and $0.44(34)$, respectively. For the second $5/2^+$ state the shell model calculations predict $0.26$, smaller than the experimental value of 0.65(15), however, the spin and parity assignment is not certain. The calculations also predict a rather small value for the $3/2^+_1$ state, $C^2S=0.30$, which is consistent with the non-observation of direct population. 
\begin{table}[h]         
  \caption{\label{tab:specfac} Experimental and shell model spectroscopic factors for the \dtsr{94,95,96} reactions. The uncertainties in the spectroscopic factors are statistical ones only. Additional reaction model uncertainties as discussed in Section~\ref{sec:ana} must be considered for absolute comparisons.
    For the odd-$N$ isotopes the ordering of the lowest predicted states are in good agreements with the experimental level schemes. For \nuc{94}{Sr} the shell model states are associated with their most probable experimental counter part. States with very low spectroscopic factors have been omitted in this table for clarity.} 
  \begin{center}       
    \begin{threeparttable}
      \begin{tabular*}{\columnwidth}{l@{\extracolsep{\fill}}rrlrr}
        \hline
        \hline 
        \multicolumn{3}{c}{experiment}&\multicolumn{3}{c}{shell model}\T\\
        $J^\pi$ & $E$ (keV) & $C^2S$ &$J^\pi$ & $E$ (keV) & $C^2S$\B\\
        \hline
        \hline
        \multicolumn{6}{c}{\dtsrf{94}{93}} \T\\
        $5/2^+$\tnote{a} &    0 & 3.37(67)   & $5/2^+$ &   0 & 4.03 \\
        $(1/2^+)$ &  213 & 0.44(34)          & $1/2^+$ & 196 & 0.87 \\
        $(3/2^+)$ &  433 &                   & $3/2^+$ & 532 & 0.30 \\
        $(5/2^+)$ & 1143 & 0.65(15)\tnote{b} & $5/2^+$ & 924 & 0.26 \\
        \hline
        \hline
        \multicolumn{6}{c}{\dtsrf{95}{94}}\T\\
        $0^+$ &    0 & 0.336(7)     & $0^+$ &    0 & 0.45 \\
        $2^+$ &  837 & 0.725(25)    & $2^+$ & 1279 & 1.14 \\
        $0^+$ & 1880 & 0.067(4)     & $0^+$ & 2315 & 0.35 \\
        $4^+$ & 2146 &              & $4^+$ & 2474 & 0    \\
        $0^+$ & 2293 & 0.105(6)     & $0^+$ & 2441 & 0.03 \\
        $3^+$ & 2414 & 0.99(3)      & $3^+$ & 1989 & 2.80 \\
        $(2,3)^+$ & 2614 & 0.21(1)  & $2^+$ & 2456 & 0.86 \\
        $(2)^+$ & 2704 & 0.45(2)    & $3^+$ & 2600 & 0.12 \\
        $(2,3)^+$ & 2921 & 0.07(1)  & $2^+$ & 3525 & 0.13 \\
        $1^+$ & 3075 & 0.058(5)     & $1^+$ & 2811 & 0.00 \\
        $1^+$ & 3172 & 0.053(5)     & $1^+$ & 3231 & 0.11 \\
        $(2,3)^+$ & 3527 & 0.079(8) & $2^+$ & 3525 & 0.13 \\
        \hline
        \hline        
        \multicolumn{6}{c}{\dtsrf{96}{95}}\T\\
        $1/2^+$\tnote{a} &    0 & 0.23(15) & $1/2^+$ &    0 & 1.57 \\
        $3/2^+$ &  352 &        & $3/2^+$ &  412 & 0.39 \\
        $5/2^+$ &  681 & 2.15(50) & $5/2^+$ &  585 & 4.80 \\
        $(1/2^+)$ & 1247 & 0.46(15)\tnote{c} & $1/2^+$ & 1833 & 0.20 \\
        \hline
        \hline
      \end{tabular*}
      \begin{tablenotes}\footnotesize
      \item[a] from Ref.~\cite{buchinger87}.
      \item[b] assuming $J^\pi = 5/2^+$
      \item[c] assuming $J^\pi = 1/2^+$
      \end{tablenotes}
    \end{threeparttable}
  \end{center}
\end{table}
In addition to the states shown in Table~\ref{tab:specfac} and Fig.~\ref{fig:level93}, the shell model calculation also predicts the $3/2_2^+$, $7/2_2^+$, $1/2_1^+$, and $7/2_2^+$ states at 695, 1040, 1230, and 1364~keV, respectively. Their spectroscopic factors are much smaller ($C^2S < 0.05$).

\subsection{Structure of \nuc{94}{Sr}}
\begin{figure}[h]
\includegraphics[width=\columnwidth]{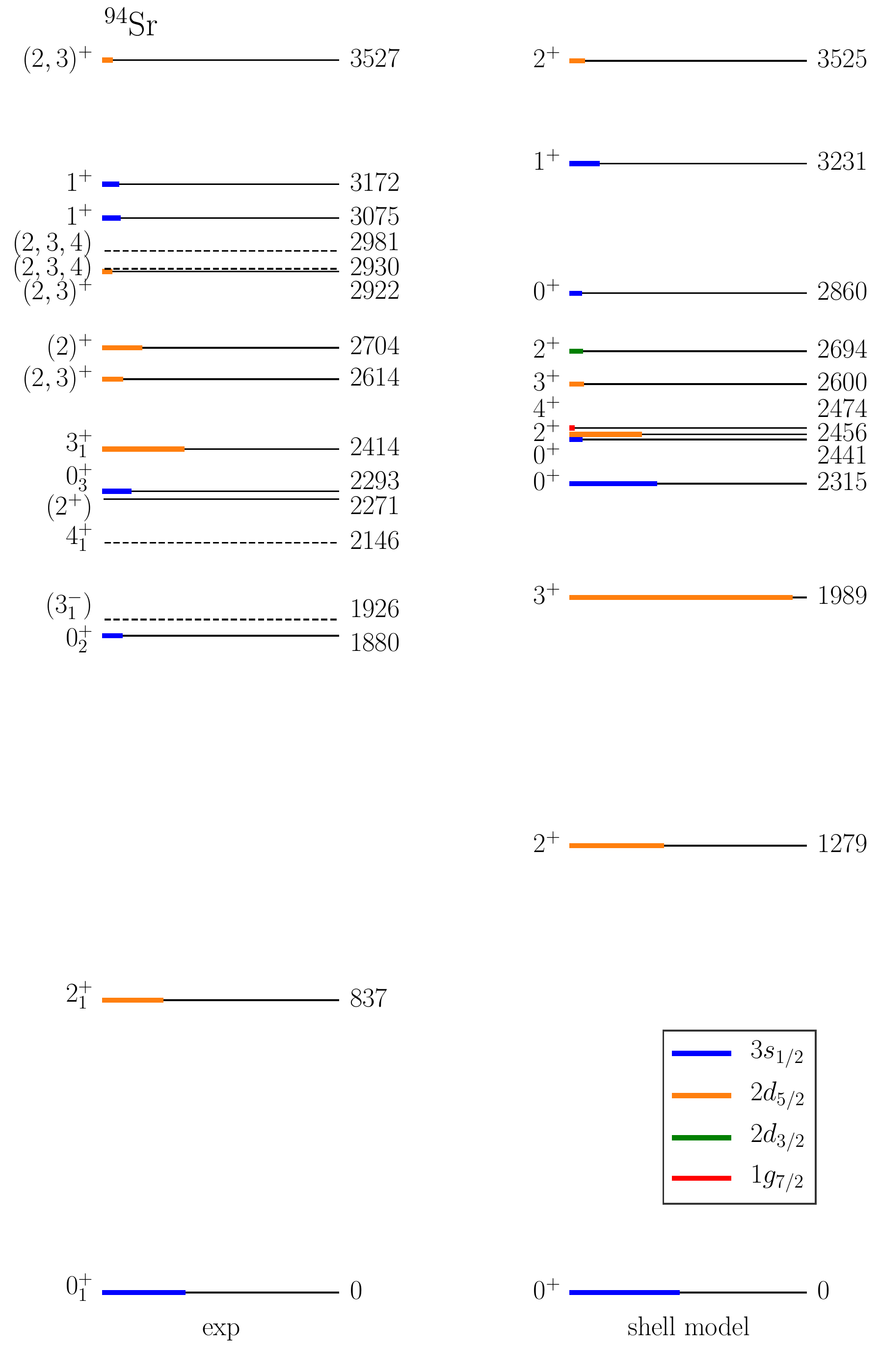}
\caption{\label{fig:level94} Level scheme of \nuc{94}{Sr}. The left side shows the experimental levels with energies in keV and the spin and parity assignments from this work. The thick lines indicate the magnitude of the experimental spectroscopic factors and the color the orbitals. The right side shows the result of shell model calculations.}
\end{figure}

The level scheme together with the spectroscopic factors obtained from the experimental cross sections is shown in Fig.~\ref{fig:level94}.
The summed spectroscopic strength for the $3s_{1/2}$ orbital amounts to $0.62(2)$ while one would naively assume that the ground state of \nuc{95}{Sr} has one neutron in $3s_{1/2}$. Taking into account the usual quenching of the spectroscopic strength by a factor of about $0.6$, this means that the observable expected $3s_{1/2}$ strength is accounted for. For the $2d_{5/2}$ orbital the summed strength is $2.53(9)$ compared to the independent particle expectation of $6$. There are several states for which the angular distribution and the spectroscopic factor could not be determined, the experimental value thus represents only a lower limit.

Also shown in Fig.~\ref{fig:level94} are the results of shell model calculations. The energy of the $2^+$ state as well as the newly discovered $0^+$ states is slightly over-predicted. The $3^+$ state on the other hand is predicted at lower energies than experimentally observed. Overall the agreement is very good, in particular, much better than for the isotope \nuc{96}{Sr}~\cite{cruz19}. Concerning the spectroscopic factors, for the ground and the $2^+_1$ state the agreement is very good. The summed spectroscopic strength for the $3s_{1/2}$ and $2d_{5/2}$ orbitals in the calculations amount to $0.97$ and $5.06$ for the states shown on the right side of Fig.~\ref{fig:level94}. The biggest discrepancy is observed for the $3^+$ state. The shell model predicts a large spectroscopic factor of $C^2S=2.80$ for this state. This can be understood in a naive independent particle model where the ground state of \nuc{95}{Sr} has a configuration with the $\nu2d_{5/2}$ orbital fully occupied, and a single neutron in the $3s_{1/2}$ orbital, $(2d_{5/2})^6(3s_{1/2})^1$. The removal of a $2d_{5/2}$ neutron results in $(2d_{5/2})^5(3s_{1/2})^1$ coupled to $J=2$ or 3.
The experimental strength for the $3^+$ state is only about one third of the calculated one. It can not be excluded that this strength of the $2d_{5/2}$ orbital is split over several states which could not be identified experimentally. This state was initially assigned as $3^-$, based on angular correlation measurements which suggest a $3\rightarrow2\rightarrow0$ cascade decay, and limit the quadrupole components for the $3\rightarrow2^+_1$ decay to less than $0.1$~\% $L=2$. Assuming a pure $M1$ transition for this decay, the measured lifetime, $\tau=5.7(23)$~ps~\cite{mach91}, corresponds to $B(M1) = 2.5(10)\cdot10^{-3}$~$\mu_\text{N}^2$, compatible but smaller than the shell model result of $B(M1) = 0.070$~$\mu_\text{N}^2$.

\begin{figure}[h]
 \includegraphics[width=\columnwidth]{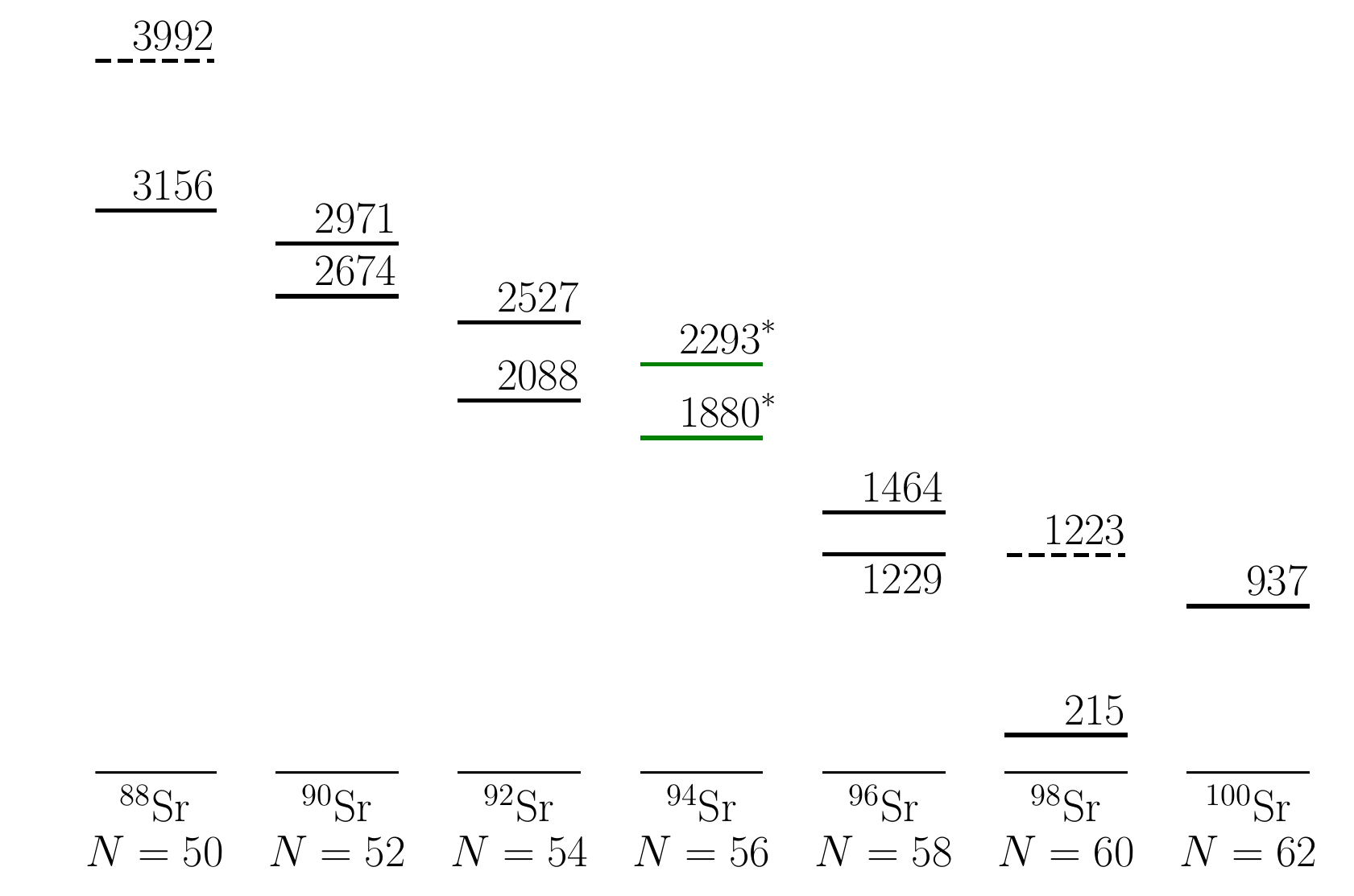}
 \caption{\label{fig:systematics} Energy systematics of the \zps\ in even-mass Sr isotopes. Levels drawn with dashed lines are tentative assignments~\cite{ragani70,kapstein78,flynn76,alquist80,jung80,schussler80,lhersonneau02}. The newly-observed \zps\ \nuc{94}{Sr} from the present experiment are marked with an asterisk.}
\end{figure}

The systematics of the \zps\ along the Sr isotopes is shown in Fig.~\ref{fig:systematics}. 
The newly discovered states at 1880 and 2293~keV fit well into this systematics and complete it up to $N=60$. The excited \zps\ smoothly come down in energy until \nuc{94}{Sr}. In \nuc{96}{Sr} a drop by about 600~keV is observed. A similar behavior is seen is the Zr isotopes where at $N=56$ the energies of the $0^+$ states have a local maximum. This, and the local maximum seen in the energies of the $2^+$ states for both Sr and Zr is an indication of the $N=56$ sub-shell closure.

The shell model calculations predict only one excited $0^+$ state with substantial spectroscopic strength, while experimentally the strength is about equally shared between the state at 1880 and 2293~keV. This resembles the situation in \nuc{96}{Sr} where the two excited \zps\ share a significant amount of spectroscopic strength and are strongly mixed. Using the same approach as was discussed for \nuc{96}{Sr} the mixing of the two excited \zps\ can be extracted as $a^2 = 1/(1+R)$ with the ratio of spectroscopic factors $R=1.55(14)$. The mixing amplitude derived from the experimental results is $a^2 = 0.39(2)$. This means that the two excited states result from the strong mixing of almost degenerate states (2042 and 2131~keV) with a mixing matrix element of $V=202(2)$~keV. The mixing of the excited \zps\ in \nuc{94}{Sr} is thus very similar to the one in \nuc{96}{Sr}. \nuc{98}{Sr} at $N=60$ however behaves completely different with very weak mixing (between the ground state and first excited $0^+$ state). Returning to the evolution of the \zps\ in the Sr isotopic chain as shown in Fig.~\ref{fig:systematics}, it is clear that the ground state shape transition at $N=60$ is accompanied by a clear structural change of the $0^+$ states. In the lighter strontium isotopes the two excited $0^+$ states strongly mix among each other, but not with the ground state, while at $N=60$ one of the excited $0^+$ states drop in energy, becomes the ground state, and a rather pure, strongly deformed state. It would be interesting to further investigate the structure of the $0^+$ states in \nuc{98}{Sr} by one-neutron transfer reactions as in the present study, however this is complicated by the highly complex structure of the \nuc{97}{Sr} ground state~\cite{buchinger87,cruz19}. 

\subsection{\nuc{95}{Sr}}
\begin{figure}[h]
\includegraphics[width=\columnwidth]{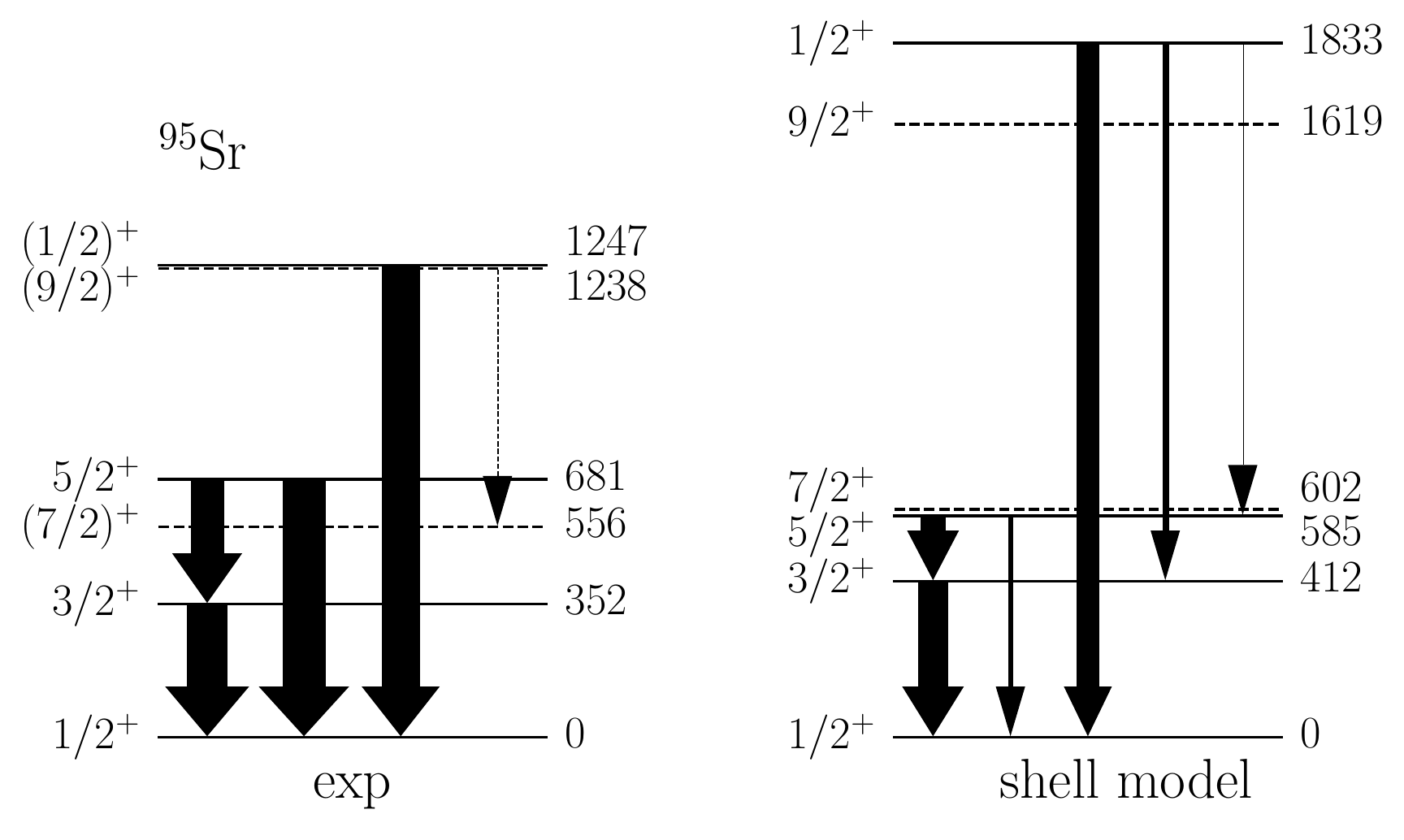}
\caption{\label{fig:level95} Proposed level scheme of \nuc{95}{Sr}. The left side shows the experimental levels with energies in keV and the spin and parity assignments from this work. The widths of the arrows are proportional to the relative intensity of the observed $\gamma$-ray transition. Dashed levels and transitions have not been observed in the present work. The right side shows the result of shell model calculations. Here the widths of the arrows indicate the branching ratios.}
\end{figure}

The level structure of \nuc{95}{Sr} is well reproduced by the shell model calculations. The level scheme is shown in Fig.~\ref{fig:level95}.
The $3/2^+$ state is not or only very weakly directly populated in the present experiment; this is in agreement with the shell model prediction that in the ground state of \nuc{96}{Sr} the $2d_{3/2}$ orbital occupation is small. While the agreement with the shell model points to the fact that \nuc{95}{Sr} is still dominated by spherical configurations, the spectroscopic factors are not expected to agree. For the inverse reaction \dpsrf{95}{96} a small spectroscopic factor of 0.19(3) was determined for the ground state transfer reaction~\cite{cruz19}. The present result of 0.23(15) is in agreement with that. In contrast the shell model predicts a spectroscopic factor of 1.57 for the ground state of \nuc{95}{Sr}. This shows that the ground state of \nuc{96}{Sr} has only little overlap with the ground state of \nuc{95}{Sr}, indicating that the transition towards strongly deformed ground state configurations at \nuc{98}{Sr} (and potentially already \nuc{97}{Sr}) is more complex then predicted. 
For the 352, and 681~keV states in \nuc{95}{Sr} the shell model calculations predict spectroscopic factors of 0.38, and 4.80. While these numbers are in qualitative agreement with the observation, the interpretation is not straightforward as the \nuc{96}{Sr} ground state does not resemble the shell model predictions.

\section{Summary}
The \dtsr{94,95,96} reactions have been studied to examine the single-particle structure of the \nuc{93-95}{Sr} isotopes toward the $N=60$ shape transition. The spin and parity assignments for excited states in \nuc{93}{Sr} have been revised. The revised level scheme is in good agreement with shell model calculations. Two excited $0^+$ states in \nuc{94}{Sr} have been observed for the first time. The similar population strength suggests strong mixing between the two $0^+$ states, resembling the situation in \nuc{96}{Sr}. Further studies, such as low-energy Coulomb excitation, could provide additional support for this interpretation. For other states, except the $3^+$ state in \nuc{94}{Sr}, good agreement with the shell model for excitation energies and spectroscopic factors is observed indicating that the structure of low-lying states in \nuc{95}{Sr} is not yet affected by the shape transition at $N=60$. The \dtsrf{96}{95} reaction is not well suited to determine the nature of states in \nuc{95}{Sr} as the ground state configuration of \nuc{96}{Sr} deviates significantly from the shell model prediction. The large overlap of the $5/2^+$ state in \nuc{95}{Sr} with the projectile ground state indicates a similar structure and suggests that this state is slightly oblate deformed. The deformation and nature of states in \nuc{95}{Sr} could be further explored by low-energy Coulomb excitation as well as lifetime measurements. Lastly, it should be mentioned that the development of shell model calculations in much larger model spaces than the ones discussed here could provide more insights in the nature of the shape transition at $N=60$.

\section{Acknowledgments} 
The efforts of the TRIUMF operations team in supplying the \nuc{94,95,96}{Sr} beam are highly appreciated. K.W. thanks A.M. Moro for his help regarding the input of the overlap function in FRESCO.
We acknowledge support from the Science and Technologies Facility Council 
(UK, grants EP/D060575/1 and ST/L005727/1), the National Science Foundation (US, grant PHY-1306297), 
the Natural Sciences and Engineering Research Council of Canada, the Canada Foundation 
for Innovation and the British Columbia Knowledge and Development Fund. TRIUMF receives federal funding via a contribution through the National Research Council Canada. K. W. acknowledges the support from the Spanish Ministerio de Econom\'ia y Competitividad RYC-2017-22007.

\bibliographystyle{bibstyle}
\bibliography{draft_used}
\end{document}